%% file: ms.tex
\documentclass[12pt,preprint]{aastex}

\newcommand{\rme}{\rm E}
\newcommand{\bdv}[1]{\mbox{\boldmath$#1$}}
\newcommand{\bpi}{\mbox{\boldmath$\pi$}}
\newcommand{\bLambda}{\mbox{\boldmath$\Lambda$}}

\def\e{{\rm E}}
\def\rel{{\rm rel}}
\def\kms{{\rm km}\,{\rm s}^{-1}}

\begin{document}

\title{\textbf{Removing the Microlensing Blending-Parallax Degeneracy Using Source Variability}}
\author{R.J.~Assef\altaffilmark{1},
A.~Gould\altaffilmark{1},\\and\\
C.~Afonso\altaffilmark{2,5},
J.N.~Albert\altaffilmark{3},
J.~Andersen\altaffilmark{4},
R.~Ansari\altaffilmark{3},
\'E.~Aubourg\altaffilmark{5},
P.~Bareyre\altaffilmark{5},
J.P.~Beaulieu\altaffilmark{6},
X.~Charlot\altaffilmark{5},
C.~Coutures\altaffilmark{5,6},
R.~Ferlet\altaffilmark{6},
P. Fouqu\'e\altaffilmark{7,8},
J.F.~Glicenstein\altaffilmark{5},
B.~Goldman\altaffilmark{2,5},
D.~Graff\altaffilmark{9,1},
M.~Gros\altaffilmark{5},
J.~Haissinski\altaffilmark{3},
C. Hamadache\altaffilmark{5},
J.~de Kat\altaffilmark{5},
L. Le Guillou\altaffilmark{10,5},
\'E.~Lesquoy\altaffilmark{5,6},
C.~Loup\altaffilmark{6},
C.~Magneville\altaffilmark{5},
J.B.~Marquette\altaffilmark{6},
\'E.~Maurice\altaffilmark{11},
A.~Maury\altaffilmark{12,8},
A.~Milsztajn \altaffilmark{5},
M.~Moniez\altaffilmark{3},
N.~Palanque-Delabrouille\altaffilmark{5},
O.~Perdereau\altaffilmark{3},
Y.R. Rahal\altaffilmark{3},
J.~Rich\altaffilmark{5},
M.~Spiro\altaffilmark{5},
P.~Tisserand\altaffilmark{5},
A.~Vidal-Madjar\altaffilmark{6},
L.~Vigroux\altaffilmark{5,6},
S.~Zylberajch\altaffilmark{5}
\\   \indent   \indent
(The EROS-2 Collaboration)\\
D.P.~Bennett\altaffilmark{13,17,18},
A.~C.~Becker\altaffilmark{14},
K.~Griest\altaffilmark{15},
T.~Vandehei\altaffilmark{15},
D.L.~Welch\altaffilmark{16}
\\   \indent   \indent
(For the MACHO Collaboration)\\
A.~Udalski\altaffilmark{19},
M.K.~Szyma{\'n}ski\altaffilmark{19},
M.~Kubiak\altaffilmark{19},
G.~Pietrzy{\'n}ski\altaffilmark{19,20},
I.~Soszy{\'n}ski\altaffilmark{19,20},
O.~Szewczyk\altaffilmark{19},
{\L}.~Wyrzykowski\altaffilmark{19,21}
\\   \indent   \indent
(The OGLE Collaboration)\\
}

\affil{
\altaffiltext{1}
{Department of Astronomy, Ohio State University,
140 W.\ 18th Ave., Columbus, OH 43210, USA;\\rjassef,gould@astronomy.ohio-state.edu}
\altaffiltext{2}
{Max-Planck-Institut f\"ur Astronomie, Koenigstuhl 17,D-69117 Heidelberg, Germany;\\afonso@mpia-hd.mpg.de,goldman@mpia-hd.mpg.de}
\altaffiltext{3}
{Laboratoire de l'Acc\'{e}l\'{e}rateur Lin\'{e}aire,
IN2P3 CNRS, Universit\'e de Paris-Sud, 91405 Orsay Cedex, France;\\albert,ansari,jhaiss,moniez,perderos,rahal@lal.in2p3.fr,marquett@iap.fr}
\altaffiltext{4}
{The Niels Bohr Institute, Copenhagen University, Juliane Maries Vej 30,
DK2100 Copenhagen, Denmark;\\JA@ASTRO.KU.DK}
\altaffiltext{5}
{CEA, DSM, DAPNIA,
Centre d'\'Etudes de Saclay, 91191 Gif-sur-Yvette Cedex, France;
\\charlot,coutures,glicens,Michel.Gros,cmv,mimile,nathalie,rich,tisseran@hep.saclay.cea.fr; eric@aubourg.net,pierre.bareyre@cdf.in2p3.fr,clarisse.hamadache@cdf.in2p3.fr,jean.dekat@antigone.cea.fr,
\\lesquoy@in2p3.fr,alain@spaceobs.com,mspiro@admin.in2p3.fr,zylberajch@wanadoo.fr}
\altaffiltext{6}
{Institut d'Astrophysique de Paris, INSU CNRS,
98~bis Boulevard Arago, 75014 Paris, France;\\beaulieu,ferlet,loup,alfred,vigroux@iap.fr}
\altaffiltext{7}
{Observatoire Midi-Pyr\'en\'ees, Laboratoire d'Astrophysique (UMR 5572),
14 av. E. Belin, 31400 Toulouse, France; pfouque@ast.obs-mip.fr}
\altaffiltext{8}
{European Southern Observatory (ESO), Casilla 19001, Santiago 19, Chile}
\altaffiltext{9}
{Division of Medical Imaging Physics, Johns Hopkins University Baltimore, MD 21287-0859, USA;\\dgraff3@jhmi.edu}
\altaffiltext{10}
{Instituut voor Sterrenkunde, Celestijnenlaan 200 B, B-3001 Leuven,Belgium;\\Laurent.LeGuillou@ster.kuleuven.be}
\altaffiltext{11}
{Observatoire de Marseille,
2 place Le Verrier, 13248 Marseille Cedex 04, France;\\eric.maurice@oamp.fr}
\altaffiltext{12}
{San Pedro de Atacama Celestial Exploration, Casilla 21, San Pedro de Atacama, Chile}
\altaffiltext{13}
{Microlensing Observations for Astrophysics (MOA) Collaboration}
\altaffiltext{14}
{Department of Astronomy, University of Washington, Box 351580, Seattle, WA 98195;\\becker@astro.washington.edu}
\altaffiltext{15}
{Physics Dept. 0319, University of California, San Diego,
La Jolla CA 92093;\\kgriest@ucsd.edu, vandehei@sbcglobal.net}
\altaffiltext{16} 
{Department of Physics \& Astronomy, McMaster University, Hamilton, Ontario
Canada L8S 4M1;\\welch@physics.mcmaster.ca}
\altaffiltext{17}
{Probing Lensing Anomalies NETwork (PLANET) Collaboration}
\altaffiltext{18}
{Department of Physics, Notre Dame University, Notre Dame, IN 46556, USA;\\bennett@nd.edu}
\altaffiltext{19}
{Warsaw University Observatory, Al.~Ujazdowskie~4, 00-478~Warszawa,
Poland;\\udalski@astrouw.edu.pl}
\altaffiltext{20}
{Universidad de Concepci{\'o}n, Departamento de Fisica,
Casilla 160--C, Concepci{\'o}n, Chile}
\altaffiltext{21} 
{Institute of Astronomy  Cambridge University,
Madingley Rd., CB3 0HA Cambridge, UK;\\wyrzykow@ast.cam.ac.uk}
}

\begin{abstract}
Microlensing event MACHO 97-SMC-1 is one of the rare microlensing
events for which the source is a variable star, simply because most
variable stars are systematically eliminated from microlensing
studies. Using observational data for this event, we show that the
intrinsic variability of a microlensed star is a powerful tool to
constrain the nature of the lens by breaking the degeneracy between
the microlens parallax and the blended light. We also present a
statistical test for discriminating the location of the lens based on
the $\chi^2$ contours of the vector $\bdv{\Lambda}$, the inverse of
the projected velocity. We find that while SMC self lensing is
somewhat favored over halo lensing, neither location can be ruled out
with good confidence.
\end{abstract}


\section{Introduction}\label{sec:intro}

Microlensing parallax measurements are a potentially powerful way to
constrain the nature of the lenses.  For typical events, the only
measured parameter that is related to the underlying physical
properties of the lens is the Einstein timescale $t_\e$.  However,
this relation is rather indirect,
\begin{equation}\label{eq:t_e}
t_\e = {\theta_\e\over \mu_\rel},
\qquad 
\theta_\e = \sqrt{\kappa M \pi_\rel}
\end{equation}
where $\pi_\rel$ and $\mu_\rel$ are the lens-source relative parallax
and proper motion, $\theta_\e$ is the angular Einstein radius, $M$ is
the mass of the lens, and $\kappa\equiv 4 G/c^2 {\rm AU}\sim
8.14\,{\rm mas}\,M_\odot^{-1}$.  That is, $t_\e$ is effectively a
combination of three parameters describing the lens and the source:
$M$, $\pi_\rel$, and $\mu_\rel$.  If one can also measure the Einstein
radius projected onto the observer plane, $\tilde r_\e$, or
equivalently the microlens parallax,
\begin{equation}\label{eq:pi_e}
\pi_\e = {{\rm AU}\over \tilde r_\e} = \sqrt{\pi_\rel\over \kappa M},
\end{equation}
then this three-fold degeneracy can be reduced by one dimension,
allowing better constraints on the mass.  Moreover, one can
disentangle the mass from the kinematic variables and so extract a
purely kinematic quantity, the projected velocity,
\begin{equation}\label{eq:v_tilde}
\tilde v = {\tilde r_\e\over t_\e} = {{\rm AU}\mu_\rel\over \pi_\rel}.
\end{equation}
This quantity is particularly useful for understanding the nature of
the lenses detected toward the Magellanic Clouds (MCs), which is
currently under debate.  The MACHO collaboration \citep{alcock00}
detected 13 to 17 such events and argued that the majority of these
were due to MACHOs making up about 20\% of the Milky Way dark halo,
whereas the EROS collaboration \citep{afonso03,tisserand05} argued
that their relative lack of such detections was consistent with all
the events being due to stars in the Milky Way disk or the MCs
themselves. All events detected so far toward the SMC have been very
bright, even though the detection of fainter events was expected. We
recognize this statistical issue but we will not address it any
further on this work.  One way to determine the nature of individual
lenses is to measure their parallax $\pi_\e$ and so their projected
velocity $\tilde v$.  \citet{boutreux96} showed that this quantity
differs greatly depending on the lens population: $\sim 50\,\kms$ for
lenses in the Galactic disk, $\sim 300\,\kms$ for halo lenses, and
$\sim 2000\,\kms$ for lenses within the MCs themselves.

However, to date it has been possible to measure microlens parallaxes
for only a handful of events (\citealt{poindexter05} and references
therein).  Measurement requires that one compare $\tilde r_\e$ to some
``standard ruler'' in the observer plane, which must therefore be of
comparable size to this quantity, i.e., of order 1 AU. While $\pi_\e$
could be routinely measured by comparing the microlensing event as
observed from the Earth and from a satellite in solar orbit
\citep{refsdal66}, the only standard ruler generally available for
ground-based observations is the Earth's orbit itself.  Hence, unless
the event takes a substantial fraction of a year (in practice $t_\e
\ga 90\,{\rm days}$), it is rarely possible to measure its parallax.

The problem is that the microlens parallax is actually a 2-dimensional
vector quantity, $\bpi_\e$, whose magnitude is $\pi_\e$ and whose
direction is that of the lens motion relative to the source.  The
component of $\bpi_\e$ parallel to the direction of the Sun at the
peak of the event, $\pi_{\e,\parallel}$, induces a distortion in the
light curve that is asymmetric with respect to the peak, while the
other component, $\pi_{\e,\perp}$, induces a symmetric distortion.
While $\pi_{\e,\parallel}$ is relatively easy to measure, even for
comparatively short events \citep{gmb94}, $\pi_{\e,\perp}$ is subject
to various degeneracies, both discrete and continuous
\citep{smith03,an04,gould04}.

The fundamental problem is that changes in any one of four of the five
parameters that describe standard microlensing also induce symmetric
distortions in the light curve, and the combination of these can often
mimic the effects of $\pi_{\e,\perp}$.  The light curve of a standard
(non-parallax) microlensing event is modeled by
\begin{equation}\label{eq:f_t}
f_k(t) = f_{{\rm s},k} A[u(t)] + f_{{\rm b},k},
\end{equation}
where the magnification $A$ is given by \citep{einst36,pac86},
\begin{equation}\label{eq:A}
A(u) = {u^2 + 2\over u\sqrt{u^2+2}},
\end{equation}
and $u$ is the source-lens separation in units of $\theta_\e$, which is
given by the Pythagorean theorem,
\begin{equation}\label{eq:u}
u = \sqrt{\beta^2 + \tau^2}
\end{equation}
in terms of $\tau$ and the impact parameter $\beta$,
\begin{equation}\label{eq:tau_beta}
\tau(t)= {t-t_0\over t_\e},\qquad \beta = u_0.
\end{equation}
Here $f_k(t)$ is the observed flux, $f_{{\rm s},k}$ is the flux of the
microlensed source, and $f_{{\rm b},k}$ is the flux from any unlensed
background light, all as observed at the $k$th observatory.
Inspection of these equations shows that changes in $u_0$, $t_\e$,
$f_{\rm s}$ and $f_{\rm b}$ all induce effects on the light curve that
are symmetric with respect to $t_0$.

If it were somehow possible to measure $f_{\rm b}$ independent of the
light curve, then the problems posed by these degeneracies would be
greatly reduced.  First, since the baseline flux, $f_{\rm base}=
f_{\rm s} + f_{\rm b}$, is generally determined very precisely,
measurement of $f_{\rm b}$ immediately yields $f_{\rm s}$.  Next,
since the peak flux $f_{\rm peak}$ is also usually well measured, one
can then also directly determine $u_0$ from $A(u_0) = (f_{\rm
peak}-f_{\rm b})/f_{\rm s}$.  This implies that only $t_\e$ and
$\pi_{\e,\perp}$ must really be simultaneously determined.  Since
changing these induces rather different light-curve distortions, it is
relatively easy to disentangle them.

To date, three methods have been proposed to determine $f_{\rm b}$:
astrometry \citep{alard95,ghosh04}, precise imaging \citep{han97}, and
manipulation of a series of images using image subtraction
\citep{gould02}.  These methods can all help, but they are
fundamentally unable to resolve blended light due to a binary
companion of the source that does not participate in the microlensing
event.  Such companions could lie at separations of order 10 AU, which
corresponds to angular separations of order 1 mas or smaller.  For the
faint sources typical of microlensing events, this is too close to be
resolved using current or foreseen instruments.

Here we propose a new method to resolve blended light.  The method
requires that the source be a regular variable, which is fairly rare
for sources in typical microlensing fields.  However, the method does
not suffer from the limitations of the other techniques: its ability
to disentangle the source from the blended light works equally well
regardless of their angular separation.  We apply this technique to
MACHO-97-SMC-1, which was a relatively long event whose source
exhibited regular variability with an amplitude of about 3\%.  We show
that this variability significantly constrains the blending and so
improves the precision of the parallax measurement.

\section{Data}\label{sec:data}

MACHO 97-SMC-1 was discovered by the MACHO collaboration
\citep{macho97} and independently by the EROS collaboration
\citep{eros98}. The OGLE collaboration also observed the event during
its late decline in its OGLE-II phase \citep{ogle297} and at baseline
in its OGLE-III phase \citep{ogle303}.

MACHO observations were mostly made at the dedicated $50''$ Great
Melbourne telescope at Mount Stromlo, Australia from June 1993 to
January 2000 with simultaneous imaging in ``$V_M$'' (4500--5900 \AA)
and ``$R_M$'' (5900--7800 \AA) passbands, but also at CTIO in the
Johnson-Cousins $R$ filter from May through November 1997.

EROS observations were made at the dedicated 1m MARLY telescope at the
European Southern Observatory at La Silla, Chile from July 1996 to
February 2003 with simultaneous imaging in ``$V_E$'' (4200--7200 \AA,
peak at 5600 \AA) and ``$I_E$'' (6200--9200 \AA, peak at 7600 \AA)
wide passbands.

OGLE observations were made at the $1.3$ meter Warsaw telescope at Las
Campanas Observatory, Chile in Cousins $I$ band from June 1997 to
November 2000 for the OGLE-II phase and from June 2001 to August 2005
for the OGLE-III phase.

Because the OGLE-III observations began after the event was over, they
cannot be used to directly constrain the microlensing event. We
incorporate them only to better understand the variability. To this
end, we fix the blending at a value similar to that of OGLE-II, since
otherwise this parameter would be completely degenerate.

\cite{eros99} showed that the source star is an intrinsic variable
with a period of 5.126 days. They fit the event simultaneously for
microlensing and intrinsic variations and used OGLE-II data to
demonstrate that some of the apparent source in the EROS images was
unrelated blended light and to better define the period of
variability. The amplitude of these oscillations is so small (around
3\% of the total flux) that the event was detected by EROS because it
managed to survive their variable-star cuts. The MACHO collaboration
did not initially detect this event, mainly because their ''main
sequence variable'' cut was still based on the LMC color-magnitude
diagram and had not been corrected for the SMC. It was rather detected
by coincidence while extracting variable stars for further study.

\section{Parametrization}\label{sec:params}

When a variable star is microlensed, the amplitude of the oscillations
will be magnified by the same factor as the star's flux, completely
independent of the amount of blended light. For example, if the star's
variability period is much shorter than the event timescale, the ratio
of the amplitude at baseline and at some magnified moment will yield a
measurement of the magnification at that stage. Since this measurement
is independent of the source's flux and of the blended light, knowing
the magnification at some point allows one to solve the $2\times 2$
system of equations formed by equation (\ref{eq:f_t}) at these two
moments, thus disentangling the source from the blended light and so
eliminating the degeneracy between the parallax and the blending.

In practice, measurements of the microlensing parameters are done by
fitting a microlensing model to the observations, eliminating the
requirement for the variability period to be much shorter than the
event timescale. In this process, the magnified variability amplitude
of the source will yield an extra constraint on the amount of blending
to fit for the magnification at each stage, thereby breaking the
forementioned degeneracy.

In this paper we fit the event to a standard microlensing model (eqs.
[\ref{eq:f_t}]-[\ref{eq:tau_beta}]) plus parallax and
combined with a variability model for the intrinsic variations of the source
star. Following the procedure outlined by \citet{an02}, we model the effects
of parallax by modifying equation (\ref{eq:tau_beta}), 
\begin{equation}\label{eq:tau_beta2}
\beta (t) = u_0 + \delta\beta,\ \  \tau (t) = \frac{t-t_0}{t_{\rme}} +
\delta\tau ,
\end{equation}
\noindent where
\begin{equation}\label{eq:db_dt}
\left[\delta\tau(t), \delta\beta(t)\right]\ =\ \left[\bdv{\pi}_{\rme}\cdot
\Delta \bdv{s}(t),
\bdv{\pi}_{\rme} \times \Delta \bdv{s}(t)\right] .
\end{equation}
\noindent and the parallax is evaluated in the geocentric frame. Here,
$\Delta\bdv{s}$ is the offset position of the Sun projected on the
plane of the sky in the geocentric frame. The two components of
$\bdv{\pi}_{\rme}$ are free parameters of the model that we fit to the
data. As in equation (\ref{eq:tau_beta}), $t_0$ is the time of the
maximum, $u_0$ is the closest angular distance between the source and
the lens in units of the angular Einstein radius, $\theta_{\rme}$, and
$t_{\rme}$ is the Einstein timescale of the event.

Parallax models are generically subject to a four-fold discrete
degeneracy \citep{an04,gould04,poindexter05}.  Two solutions are
related by the ``constant acceleration'' degeneracy \citep{smith03},
which in the geocentric frame sends $u_0\rightarrow -u_0$ and changes
other parameters relatively little.  The two other solutions are
related by the ``jerk-parallax'' degeneracy, in which the jerk of the
Earth's motion can masquerade as a parallax effect \citep{gould04}.
\citet{eros98} found both constant-acceleration solutions, although
they did not express them in exactly the terms presented here.

\begin{figure}
\plotone{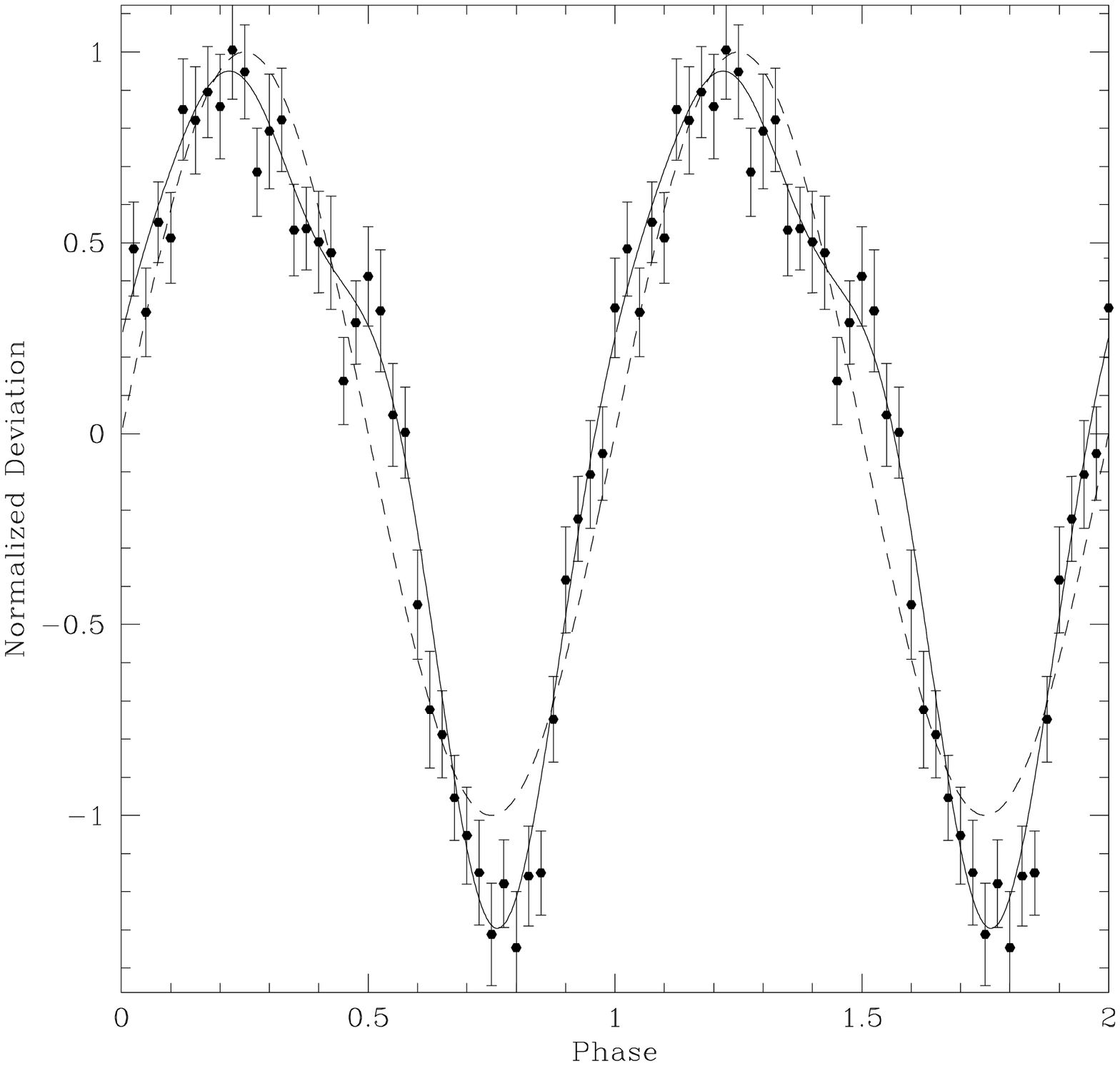}
\caption{Variability light curve of MACHO 97-SMC-1, derived from the
residuals of the fit of equation (\ref{eq:f_t_2}) with $G(x) =
\sin(x)$. The solid curve shows the variability model used in this
analysis, which correspond to a fourth-order Fourier expansion fit of
the residuals, while the dashed line shows a sine function, the model
used by \citet{eros98}.}\label{fg:light_curve}
\end{figure}

To describe the intrinsic variation of the star, we will begin by
following \cite{eros98}, who modeled the variability as a sine
function, leaving as free parameters its angular frequency $\Omega$,
its phase $\phi$, and its amplitude $\epsilon_k$ as measured at each
observatory.  Hence, the light curve is given by
\begin{equation}\label{eq:f_t_2}
f_k (t) = f_{{\rm s}, k}\ A(t)\ [1\ +\ \epsilon_k G{(\Omega t +
\phi})]\ +\ f_{{\rm b},k}\ ,\ \ G(x) \equiv \sin{x} .
\end{equation}
Our approach is very similar, but with one minor modification.  After
we initially fit the light curve to equation (\ref{eq:f_t_2}), we fit
the residuals to a fourth-order Fourier expansion to derive a new
$G(x)$, which differs modestly from a sine function (see
Fig. \ref{fg:light_curve}).

\citet{udalski97} suggested that the light curve oscillations may be
due to ellipsoidal variability of the source.  If so, this would imply
that the light curve should be subject to ``xallarap'' (binary-source
motion) as well as parallax (Earth-motion) distortions.  We will
examine this possibility more closely in Appendix A, but for the
present we simply treat these oscillations as an empirical fact of
unknown origin.

\section{Results}\label{sec:results}

Using the model described in \S~\ref{sec:params}, we fit
simultaneously the data from all the observatory/filter combinations
(7 in total). The $\chi^2$ of the best fit, in total and by
observatory/filter, is shown in Table \ref{tab:chi_2} for the positive
and negative $u_0$ solutions. It should be noted that the errors are
scaled (with scaling factors 1.92, 2.65, 1, 0.79, 1.1, 0.75 and 1
respectively for each observatory as listed in Table \ref{tab:chi_2})
so that the best fit $\chi^2$ is approximately equal to the number of
data points in each observatory/filter. Also, the best fit parameters
are listed in Table \ref{tab:best_fit} with their respective errors
for both solutions. We searched for the jerk-parallax solutions
according to the prescription of \cite{park04}. We indeed found two
additional minima on the $\chi^2$ surface, but these are excluded at
the 12-$\sigma$ and 28-$\sigma$ levels respectively, and so they will
not be further considered.

\begin{figure}
\plottwo{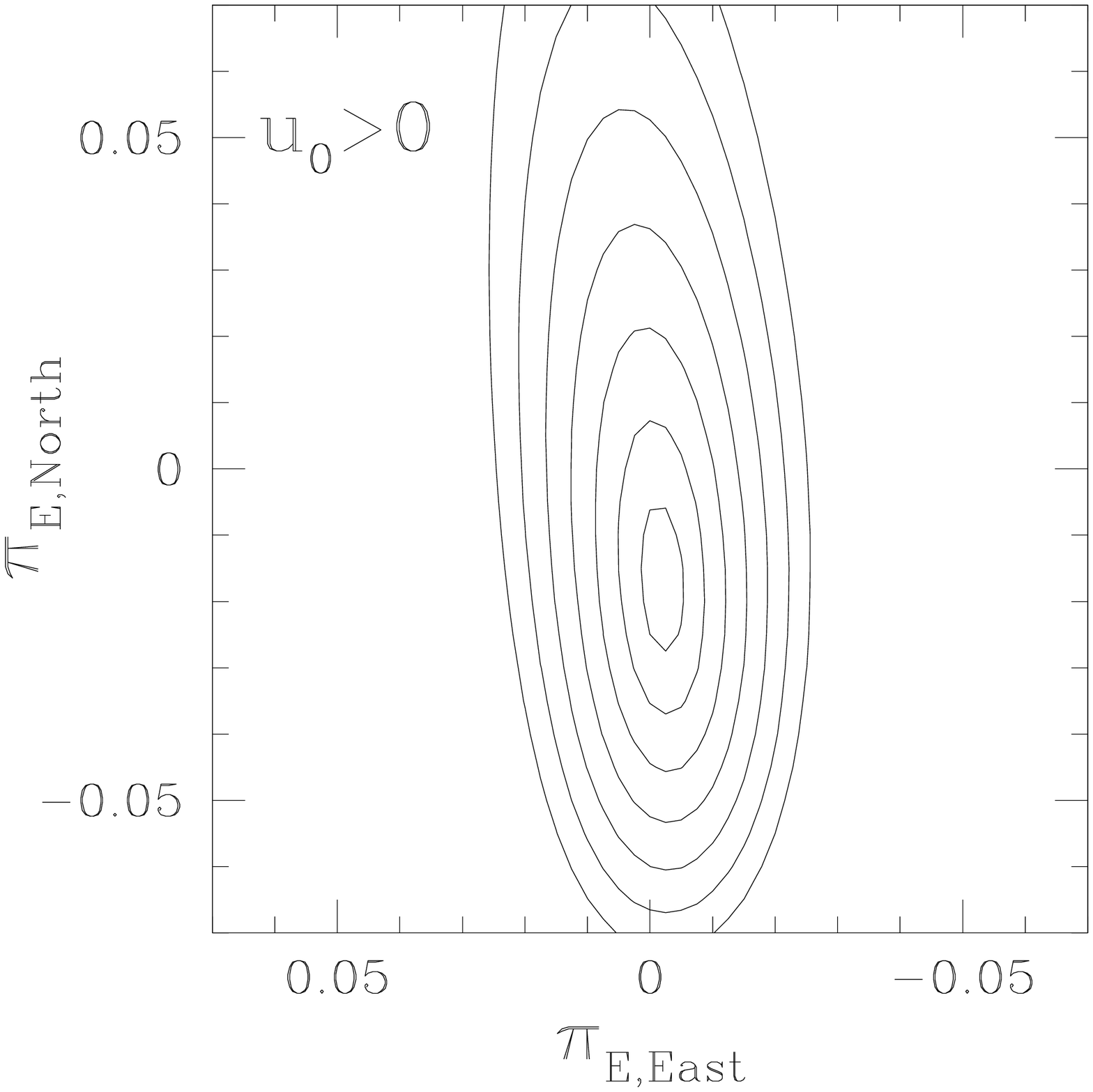}{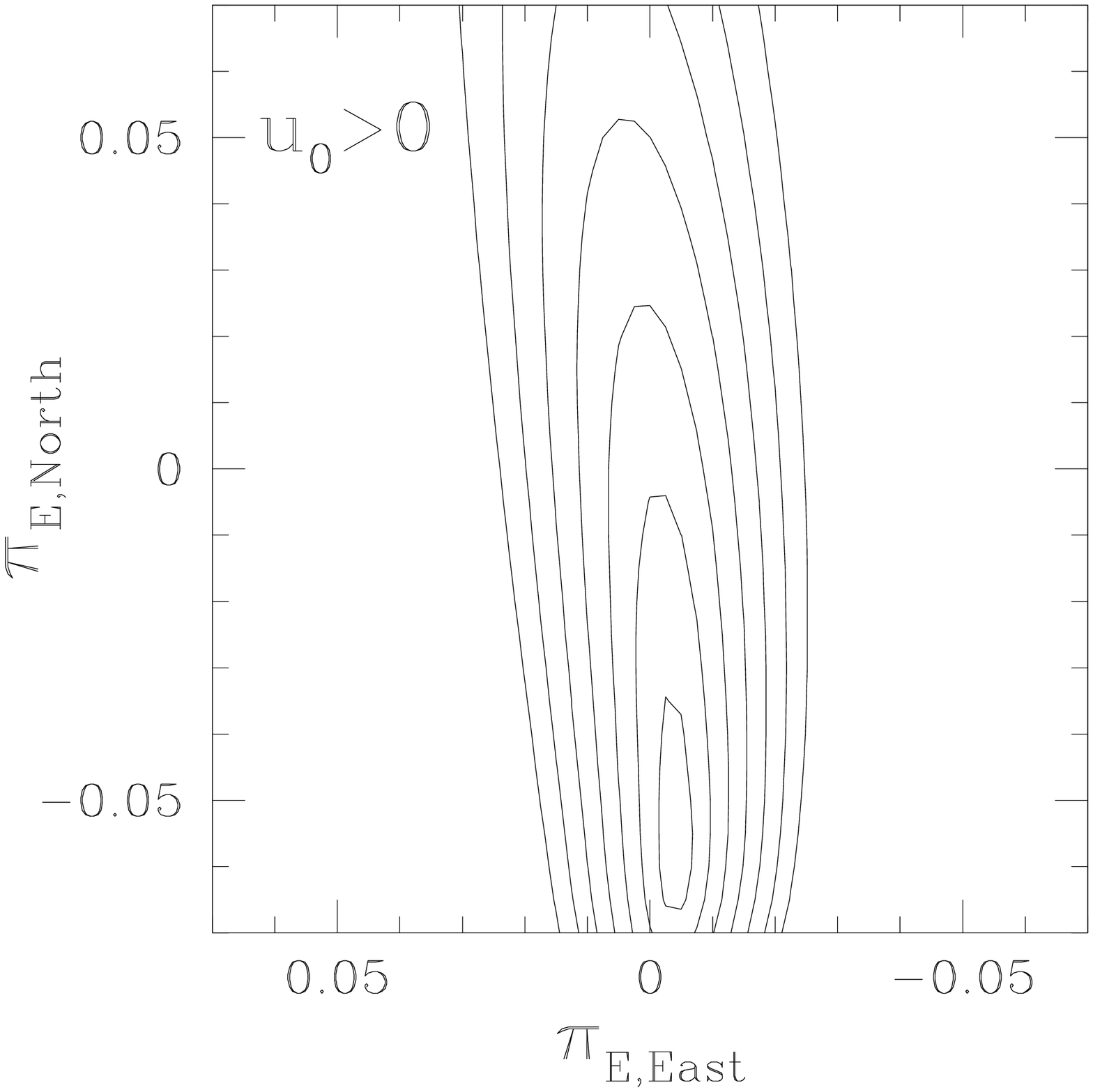}
\plottwo{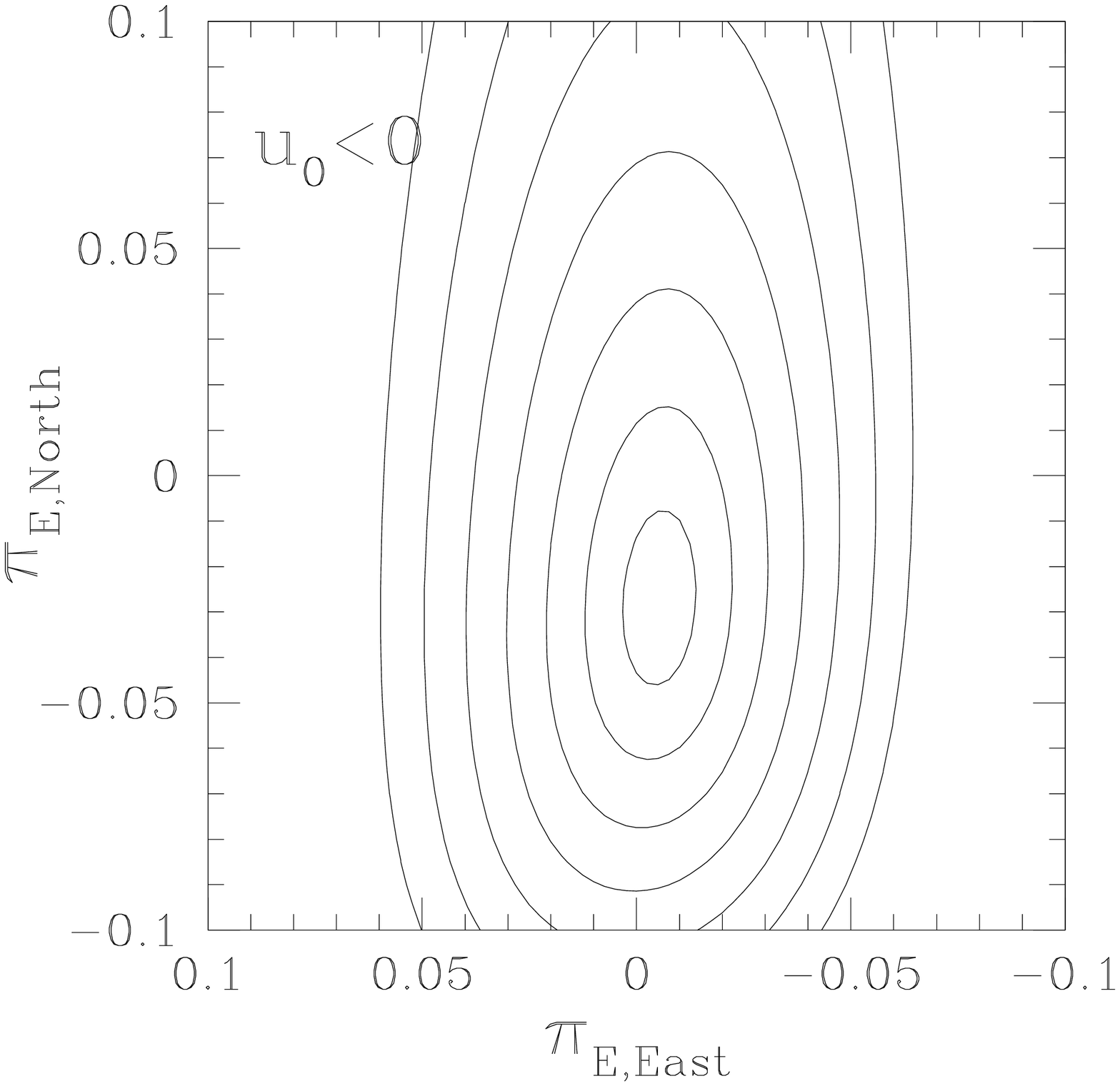}{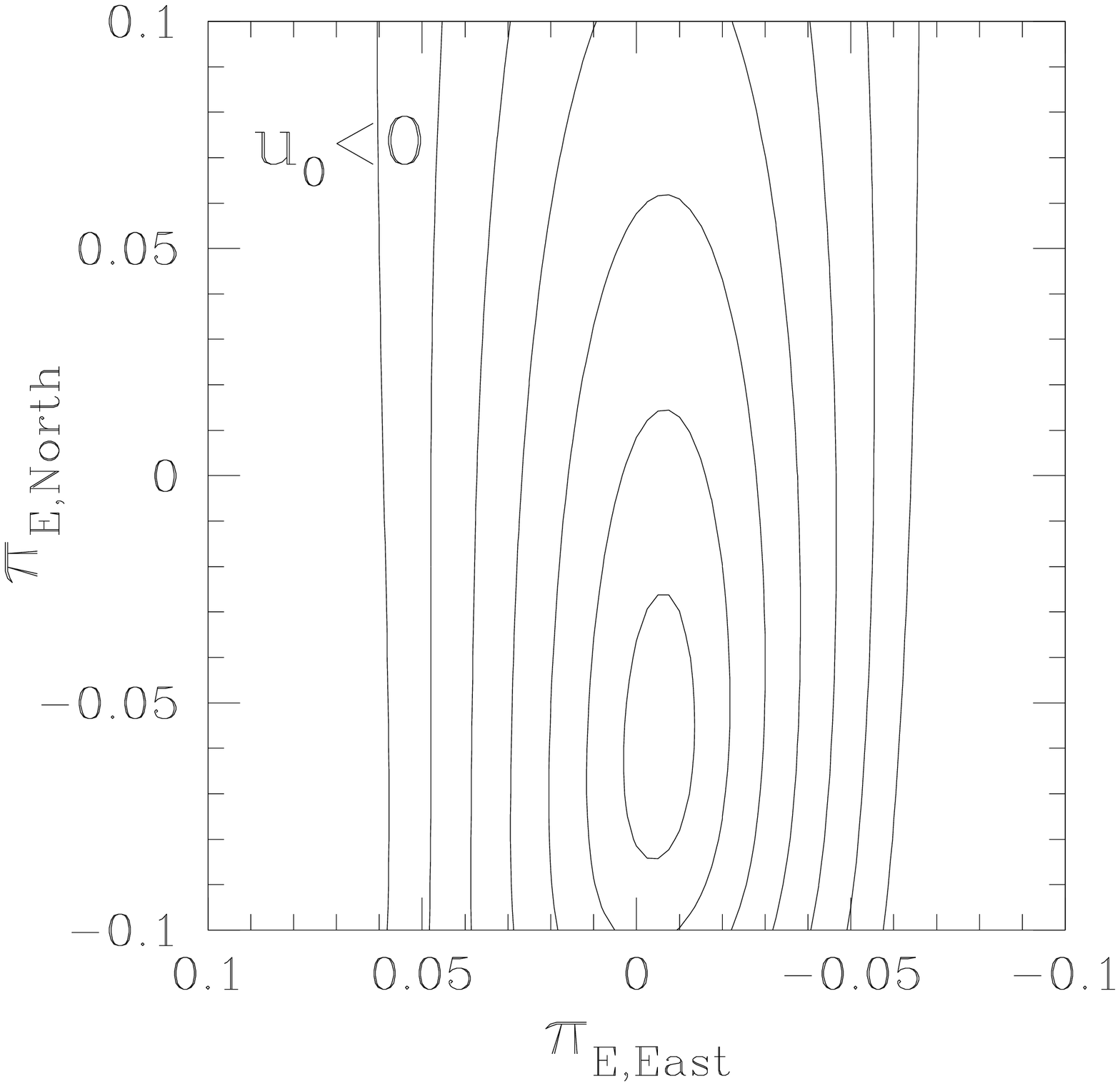}
\caption{$\Delta\chi^2$ contours (1,4,9,16,25,36,49) of the parallax
parameters for positive (top) and negative (bottom) $u_0$. The plots
on the left show these contours for the model described in
\S~\ref{sec:params}, a standard microlensing model with parallax and
taking into account the variability of the source, while the ones on
the right shows the parallax $\chi^2$ contours for the same model but
with the variability removed from both the data and the model. The
contours on the right are strongly stretched in the direction of
$\pi_{\rme,\perp}$ (which coincides almost exactly with
$\pi_{\rme,N}$) compared to the ones on the left, showing that the
variability of the source breaks the degeneracy between
$\pi_{\rme,\perp}$ and $f_{{\rm b},k}$, allowing to constrain much
better the parallax.}\label{fg:cont_var_novar}
\end{figure}

Parallax parameters are of special interest to constrain the lens
projected velocity and mass. Their best fit values show that there is
very little, if any, parallax effect. In Figure
\ref{fg:cont_var_novar}, we show $\chi^2$ contours for both ($\pm
u_0$) solutions. Although the $\chi^2$ of the best fits do not differ
significantly, the errors are a factor 2-3 smaller for positive
$u_0$. In each case, the best fit variability period for the star is
$5.1252\pm0.0001$ days, compatible with, but much more precise than,
the ones found by \cite{eros98} and \cite{udalski97}.

\citet{eros98} raised the question of whether it is the microlensed
source that is varying or it is the blended light. Their data alone
permitted them to discriminate against the second option just at the
2.5-$\sigma$ level. By including the data from all 7 observatories we
find that the variable blended-light model is ruled out at the
7-$\sigma$ level, so it will not be further considered.

In order to evaluate the role of the intrinsic source variations in
constraining the blended flux, and so $\pi_{\rme,\perp}$, we remove
the best-fit variation from the data and then refit these adjusted
data using a standard microlensing model with parallax but with no
oscillations. The $\chi^2$ and the best fit parameters are summarized
in Tables \ref{tab:chi_2} and \ref{tab:best_fit} respectively, for the
$\pm u_0$ solutions. Note that the best fit for the parameters that
are correlated with $\pi_{\rme,\perp}$, namely $u_0$, $t_E$, $f_{{\rm
s}, k}$ and $f_{{\rm b},k}$, have changed somewhat compared with the
variability-model best fit, but the main effect is that their errors
have increased by a factor of 2-3 while the $\chi^2$ of the best fit
is virtually the same.  This is reflected in the difference between
the left and right contour plots of Figure \ref{fg:cont_var_novar},
which respectively do and do not take variability into account. The
contours on the right are strongly stretched in the direction of
$\pi_{\rme,\perp}$ (note that $\pi_{\rme,\parallel}$ points just
$2^{\circ}\hskip-2pt .2$ north of west, so $\pi_{\rme,\perp}$ is
virtually the same as $\pi_{\rme,N}$). This demonstrates that if the
source star is a regular variable, the degeneracy between
$\pi_{\rme,\perp}$ and $f_{\rm b}$ can be broken and their values can
be better constrained.

\citet{wyrzy06} searched the entire OGLE-III database up through the
2004 season for microlensing events with variable baselines. Out of
about 1400 events, there were 21 with periodic variable baselines and
111 that showed a baseline with irregular variability.  The latter are
not well suited for this method, in part because apparent
"microlensing events" on irregular variables can in principle be just
a manifestation of their variability and in part because the amplitude
of variability during the event cannot be precisely determined from
the baseline variability. Naively, one would expect that half of the
periodic ones were due to a variable source and half due to variable
blend, which translates into an expectation of a little less than 1
regular variable source microlensing event for every 100 events
detected.  When considering also the probability of detecting parallax
asymmetries on the light curve, the number of events to which the
method presented in this paper is applicable is somewhat smaller. So,
even though events like MACHO 97-SMC-1 are rare, there should be a
fair number of cases for which the analysis described in this paper
would prove useful.

\section{Lens Location}\label{sec:location}

\begin{figure}
\plottwo{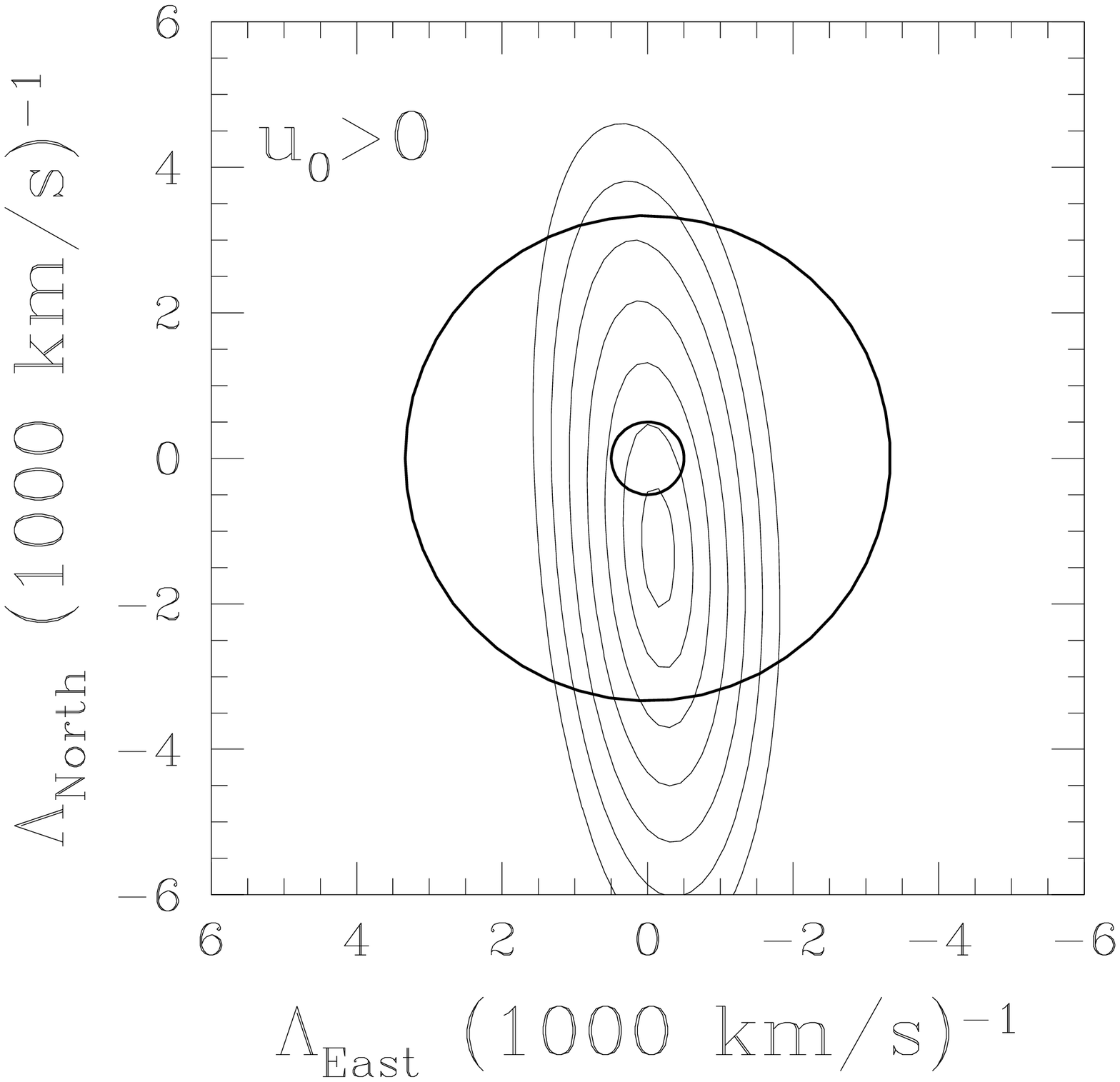}{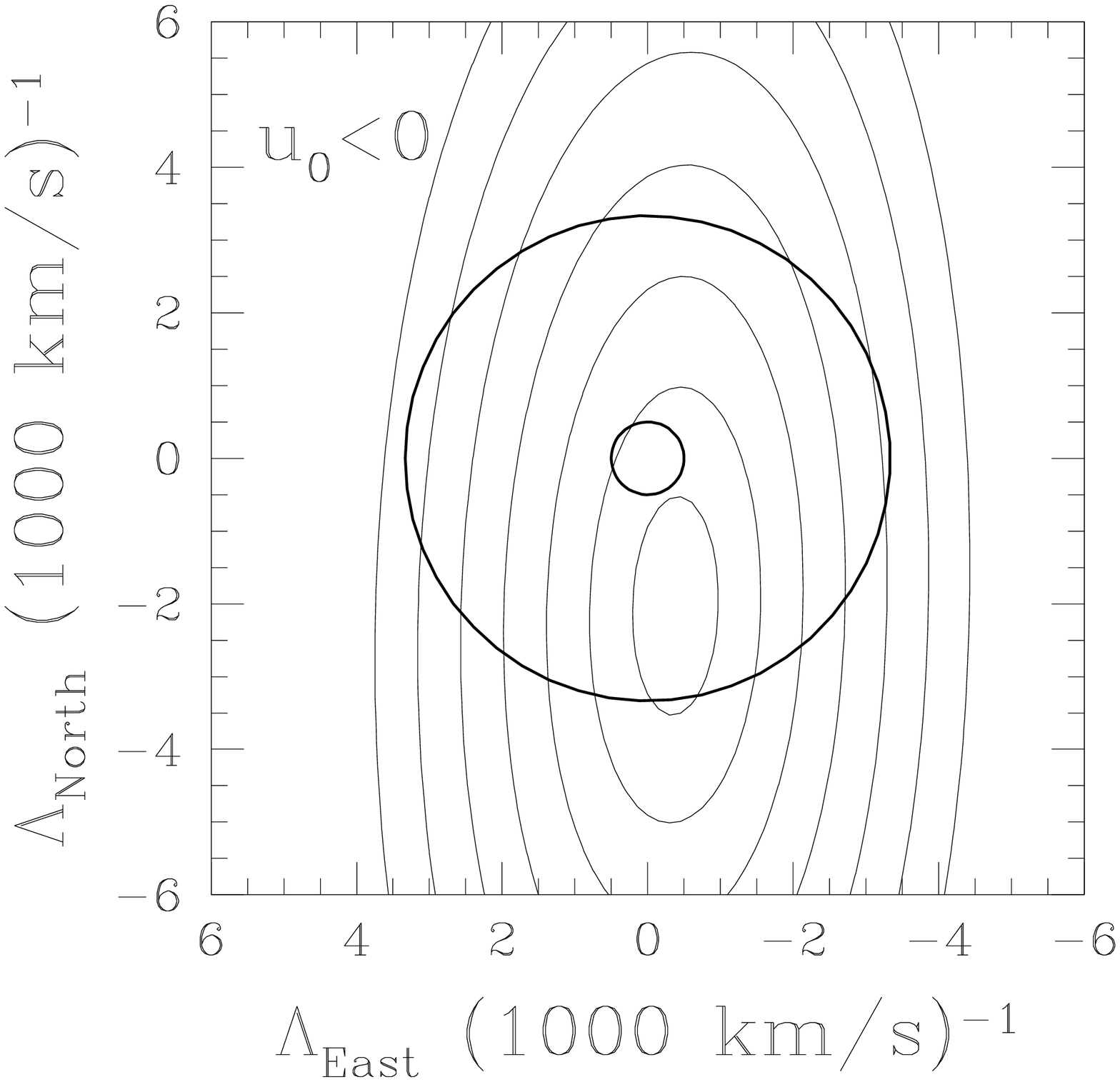}
\caption{$\Delta\chi^2$ contour (1,4,9,16,25,36,49) plots for
$\bLambda$, the inverse of the projected velocity (in $\kms$), for
both $\pm u_0$ solutions. The small circle in the middle shows a
projected velocity (in the frame of the Sun) of $2000\ \kms$ while the
other circle shows $300\ \kms$, typical projected velocities for SMC
and halo lenses, respectively. }\label{fg:lambda}
\end{figure}

A better constraint on the parallax parameters also gives a better
constraint on the lens location, since the projected velocity of the
lens depends directly on them (eqs. [\ref{eq:pi_e}] and
[\ref{eq:v_tilde}]). Figure \ref{fg:lambda} shows the $\chi^2$
contours of the vector $\bLambda$,
\begin{equation}\label{eq:Lambda}
\bLambda\ \equiv \frac{\tilde{\bdv{v}}}{\tilde{v}^2}\ =\ \frac{\bpi_{\rme}\ t_{\rme}}{\rm AU},
\end{equation}
\noindent whose direction is the same as the projected velocity but
with the inverse of its magnitude. We choose this vector instead of
the projected velocity because it is well behaved at high velocities
and makes the contours clearer. Indeed if the best fit for $t_{\rme}$
were independent of $\bdv{\pi}_{\rme}$, the $\bLambda$ contours would
differ from the $\bpi_{\rme}$ contours just by a scale factor. Note,
however, that these $\chi^2$ contours are made in the frame of the
Sun, rather than in the geocentric frame in which we conducted the
parallax analysis. That is, we convert
\begin{equation}\label{eq:v_hel}
\tilde{\bdv{v}}_{hel}\ =\ \tilde{\bdv{v}}_{geo}\ +\ \bdv{v}_{\oplus,\perp} ,
\end{equation}
\noindent where $\bdv{v}_{\oplus,\perp} = (29.75, 3.10)\ \kms$ is the
offset between the two frames. The inner circle shows $\tilde{v} =
2000\ \kms$ while the outer one shows $\tilde{v} = 300\ \kms$.

There are three possible locations for a lens that magnifies a star in
the SMC: it is located either in the Milky Way disk, the Milky Way
halo (halo lensing) or in the SMC itself (SMC self lensing). However,
a disk location is immediately ruled out by the high projected
velocity. At first sight, since the minimum $\chi^2$ is located around
a projected velocity of $1000\ \kms$ for both $u_0$ solutions, SMC
self lensing appears more probable, because this would seem to be a
too high velocity for a halo lens. To quantitatively discriminate
between these two possibilities, we compare the mean maximum
likelihood parameter, $\langle \exp[-\Delta \chi^2/2] \rangle$, for
halo and SMC lenses based on the $\chi^2$ contours of their $\bLambda$
vectors and their respective velocity distributions. We consider SMC
self lensing and halo lensing in turn.

\subsection{SMC Self Lensing}\label{ssec:self_lens}

We begin by assuming that SMC lenses are Gaussian-distributed, namely,
that they have a projected velocity probability distribution given by
\begin{equation}\label{eq:p_v}
P(\tilde{v})\ d\tilde{v}_x\ d\tilde{v}_y\ \propto\
e^{-(\tilde{v}/\sigma)^2/2}\ \tilde{v}\ d\tilde{v}_{\rm North}\
d\tilde{v}_{\rm East} ,
\end{equation}
\noindent where $\sigma$ is the velocity dispersion of the
lenses. Then, the mean likelihood parameter is
\begin{equation}\label{eq:mean_like_smc}
\langle e^{-\Delta \chi^2/2} \rangle\ =\ \int e^{-\Delta
\chi^2(\tilde{\bdv{v}})/2}\ e^{-(\tilde{v}/\sigma)^2/2}\ \tilde{v}\
d\tilde{v}_{\rm North}\ d\tilde{v}_{\rm East} \bigg/ \int
e^{-(\tilde{v}/\sigma)^2/2}\ \tilde{v}\ d\tilde{v}_{\rm North}\
d\tilde{v}_{\rm East}.
\end{equation}
The typical projected velocity of a SMC lens is not very clear but
should be of the order of $\tilde{v} \gtrsim 1000\ \kms$, so we
consider two different estimates for the velocity dispersion: $\sigma
= 1000/\sqrt{2}\ \kms$ and $\sigma = 3000/\sqrt{2}\ \kms$. As
mentioned above, the $\chi^2$ calculation is more stable in $\bLambda$
space, so we evaluate equation (\ref{eq:mean_like_smc}) in this space:
\begin{equation}\label{eq:mean_like_smc_lambda}
\langle e^{-\Delta \chi^2/2} \rangle\ =\ \int e^{-\Delta
\chi^2(\bLambda)/2}\ e^{-(\tilde{v}/\sigma)^2/2}\ \Lambda^{-5}\
d\Lambda_{\rm North}\ d\Lambda_{\rm East} \bigg/\int
e^{-(\tilde{v}/\sigma)^2/2}\ \Lambda^{-5}\ d\Lambda_{\rm North}\
d\Lambda_{\rm East} ,
\end{equation}
\noindent where $\Delta \chi^2(\bLambda)$ is taken directly from the
data of the contour plots of Figure \ref{fg:lambda}.

\subsection{Halo Lensing}\label{ssec:halo_lens}
The calculation of the mean likelihood parameter for halo lensing is
very similar to the one performed for SMC self lensing in the previous
section, but we must allow for two considerations that were implicitly
neglected before: the spatial distribution of the lenses and the
effect of the Sun's and the source's motions on the projected velocity
of the lens. We assume that halo lenses are distributed in an
isothermal sphere,
\begin{equation}\label{eq:rho}
\rho(r)\ =\ \frac{1}{a_0^2 + r^2} ,
\end{equation}
where $r$ is the lens Galactocentric distance and where we adopt $a_0
= 5\ \rm{kpc}$ for the core radius and assume that the Sun is at $R_0
= 7.6\ \rm{kpc}$. The velocity probability distribution is then
weighted by the density of lenses at each distance and by the size of
its respective Einstein ring, and integrated over all possible
distances, giving a mean likelihood parameter of
\begin{equation}\label{eq:mean_like_halo}
\langle e^{-\Delta \chi^2/2} \rangle\ =\ \int \exp{[-\Delta
\chi^2(\bLambda)/2]}\ g(\bLambda,\, D_L)\ d\, D_L\ d\Lambda_{\rm
North}\ d\Lambda_{\rm East} \bigg/\int g(\bLambda,\, D_L)\ d\, D_L\
d\Lambda_{\rm North}\ d\Lambda_{\rm East} ,
\end{equation}
\noindent where 
\begin{equation}\label{eq:g_l}
g(\bLambda,\, D_L)\ =\
\frac{\exp{[-(\tilde{\bdv{v}}_L/\sigma(D_L))^2/2]}}{2 \pi
[\sigma(D_L)]^2}\ \Lambda^{-5} D_{LS}\ \rho(D_L)\ \sqrt{\frac{D_L
D_{LS}}{D_S}}.
\end{equation}
The velocity dispersion is now given by
\begin{equation}
\sigma(D_L)\ =\ \frac{v_{rot}}{\sqrt{2}}\ \frac{D_S}{D_{LS}} ,
\end{equation}
\noindent with $v_{rot} = 220\ \kms$. Here, $D_L$ and $D_S = 60\
\rm{kpc}$ are the distances to the lens and the source, while $D_{LS}
\equiv D_S - D_L$. The extra $\sigma^{-2}$ factor in equation
(\ref{eq:g_l}) is due to the fact that we cannot ignore the
normalization of each Gaussian in this case. Note that in equation
(\ref{eq:g_l}), $\tilde{\bdv{v}}_L$ is regarded as an implicit
function of $\bLambda$, which must be made explicit in order to
quantitatively evaluate equation (\ref{eq:mean_like_halo}). Also note
that there is an extra factor of $D_{LS}$ on equation (\ref{eq:g_l}),
since the probability should be weighted by $v$ and not by
$\tilde{v}$, as was done in the previous section.

The second consideration we must take into account for halo lensing is
the effect of the motion of the Sun and the source on the projected
velocity of the lens. In equation (\ref{eq:g_l}) we are assuming that
the velocity distribution of the lenses is isotropic in the Galactic
frame but we must also include the velocity offset induced by the
motion of the Sun and the SMC with respect to the Galaxy. The three
dimensional velocity of the Sun in $UVW$ coordinates in the Galactic
frame was measured by \citet{hogg05} to be
\begin{equation}\label{eq:vel_sun}
\bdv{v}_{\odot}\ =\ (10.1,\ 224,\ 6.7)\ \kms ,
\end{equation}
\noindent where $U$ points from the Sun to the Galactic center, $V$
points towards the Sun's Galactic rotation and $W$ points towards the
Galactic north pole. The velocity of the SMC is not well measured, but
since it is most probably orbiting the LMC, it should have a very
similar three dimensional velocity. Hence, we adopt the three
dimensional velocity of the LMC for the SMC, which was measured by
\citet{kalli05} to be
\begin{equation}\label{eq:vel_smc}
\bdv{v}_{\rm{SMC}}\ =\ (-86,\ -268,\ 252)\ \kms ,
\end{equation}
\noindent in the same coordinates system. Projected on celestial north
and east coordinates, these velocities are
\begin{eqnarray}\label{eq:celest_vel}
(\bdv{v}_{\odot,\rm N},\ \bdv{v}_{\odot,\rm E})\ =\ (119,\ -133)\\ 
(\bdv{v}_{\rm{SMC},\rm N},\ \bdv{v}_{\rm{SMC},\rm E})\ =\ (-296,\  229) .
\end{eqnarray}
The projected velocity of the lens relative to the observer-source
line of sight is then
\begin{equation}\label{eq:corr_proj_vel}
\tilde{\bdv{v}}\ =\ \tilde{\bdv{v}}_L\ -\ \tilde{\bdv{v}}_{0}(D_L) ,
\end{equation}
\noindent with
\begin{equation}\label{eq:v_0_z}
\tilde{\bdv{v}}_{0}(D_L)\ \equiv\ \bdv{v}_{\odot}\ +\ \frac{D_L}{D_{LS}}\bdv{v}_{\rm{SMC}} .  
\end{equation}
Solving equation (\ref{eq:corr_proj_vel}) for $\tilde{v}_L$ and
substituting into equation (\ref{eq:g_l}), we find that for the
positive $u_0$ solution, the ratios of the mean likelihood parameters
of the SMC to the halo are 7 and 3 for the larger and smaller velocity
dispersion respectively, while for the negative $u_0$ solution they
are 3 and 2 respectively. This means that while SMC self lensing is
somewhat favored, neither of the locations can be ruled out with good
confidence.

\begin{figure}
\plottwo{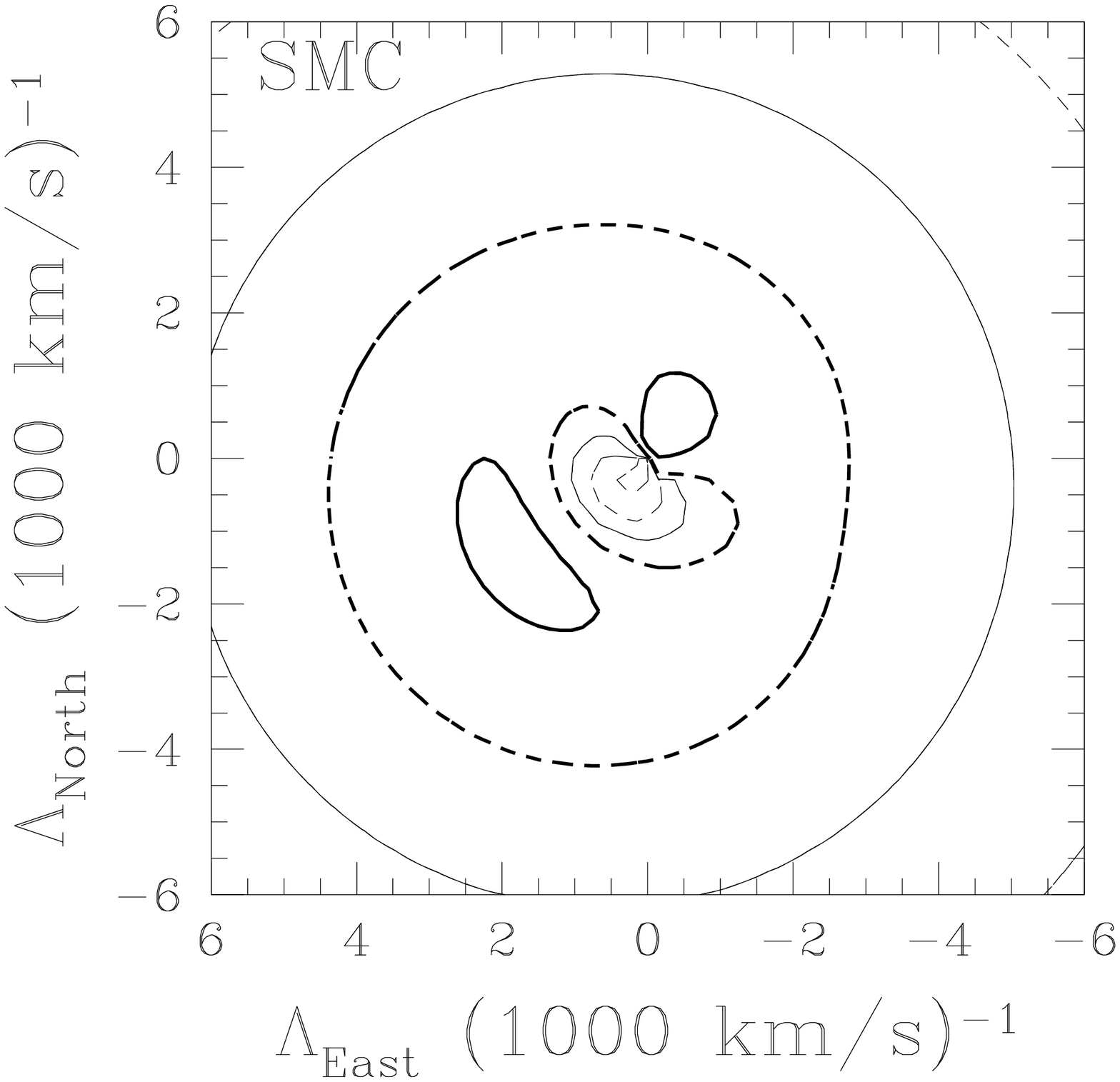}{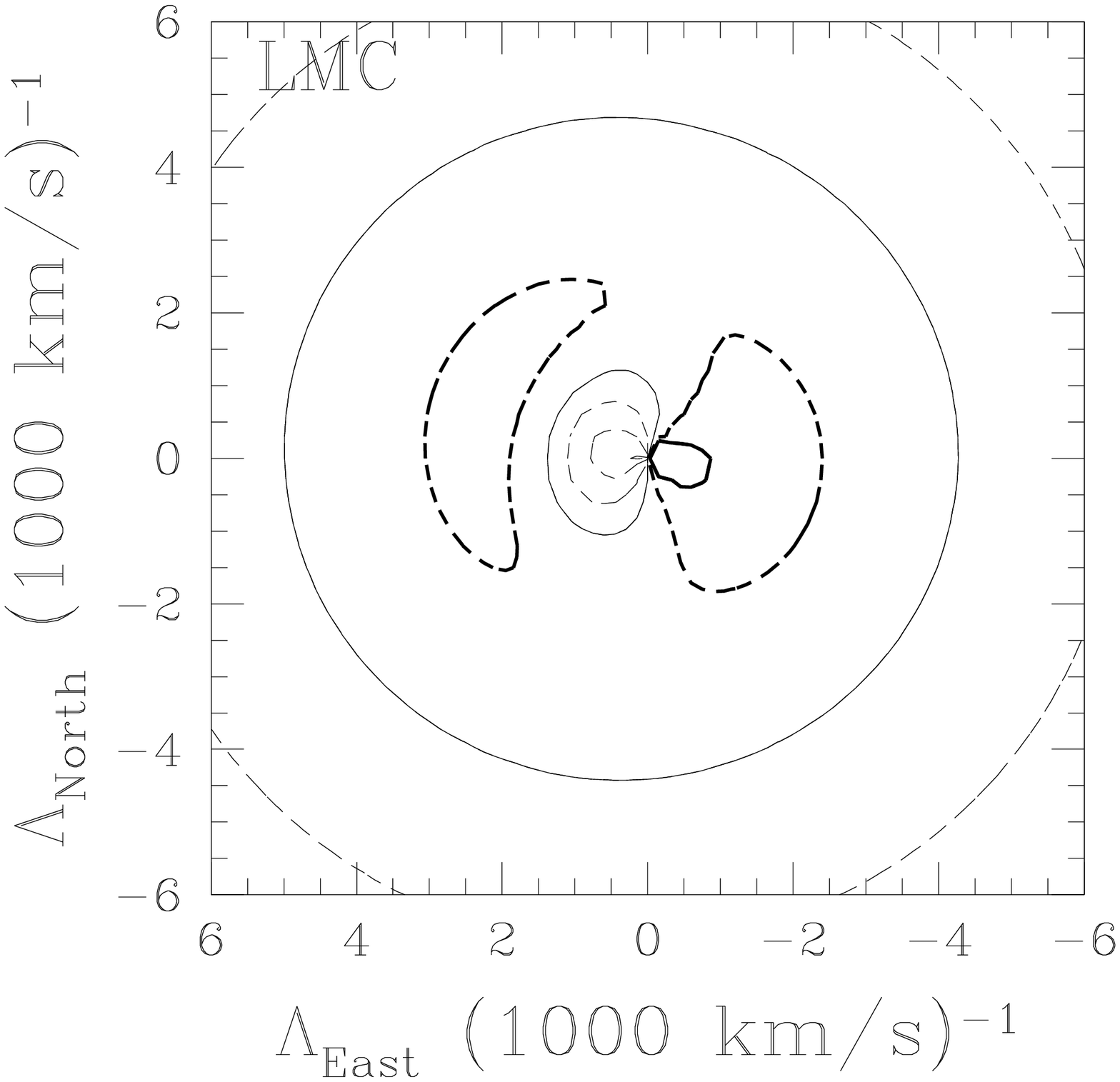}
\caption{Inverse projected velocity $\bLambda$ probability
distribution of halo lenses causing microlensing events toward the SMC
(left) and the LMC (right). The velocity offset discussed in
\S~\ref{ssec:halo_lens} makes the distributions highly asymmetric,
contrary to the case for SMC self lensing. The contour levels are
offset from one another by a factor of 5, with order, highest to
lowest, bold solid, bold dashed, thin solid and thin dashed (of which
there are several).}\label{fg:SMC_LMC_weight}
\end{figure}

In the left panel of Figure \ref{fg:SMC_LMC_weight} we show the
$\bLambda$ probability distribution for halo lenses magnifying stars
in the SMC, integrated over all $D_L$. There are two probability peaks
with similar $\Lambda$ (or speed) but roughly opposite direction. This
is due to the velocity offset, which flips sign as $D_L$
increases. Also, for completeness of the statistical method proposed
in this paper, we show in the right panel of Figure
\ref{fg:SMC_LMC_weight} the same probability distribution but for halo
lenses causing microlensing events toward the LMC, in which an
asymmetric probability peak can also be seen.

It should be noted that throughout this statistical test, we have
assumed that there is equal prior probability that a microlensing
event is caused either by self lensing or by halo lensing. There is no
evidence for assuming something different, so the ratio of the mean
likelihood parameters is actually the ratio of the probabilities of
having this event caused by self lensing and by halo lensing.

\section{EROS Fixed Blending}\label{sec:eros_fix_blend}

OGLE observations showed the presence of an optical companion to the
source of MACHO 97-SMC-1 that was unresolved by
EROS. \citet{udalski97} estimated the light contribution of the
companion to be 23\%--28\% of the total $I$ band flux. This compares
with $28\pm 12\ \%$ for the fits shown in Table \ref{tab:best_fit},
which tends to confirm that the OGLE-resolved source is the true
microlensed source (although it remains possible in principle that
there is a small component of blended light within this source). In
this section we therefore assume that the blended flux in the EROS
$I_E$ band is exactly 24\% and repeat the analysis of
\S~\ref{sec:results}. Such a strong constraint should remove the
degeneracy between $f_{\rm b}$ and $\pi_{\rme,\perp}$, perhaps leading
to a better determination of the lens location.

\begin{figure}
\plottwo{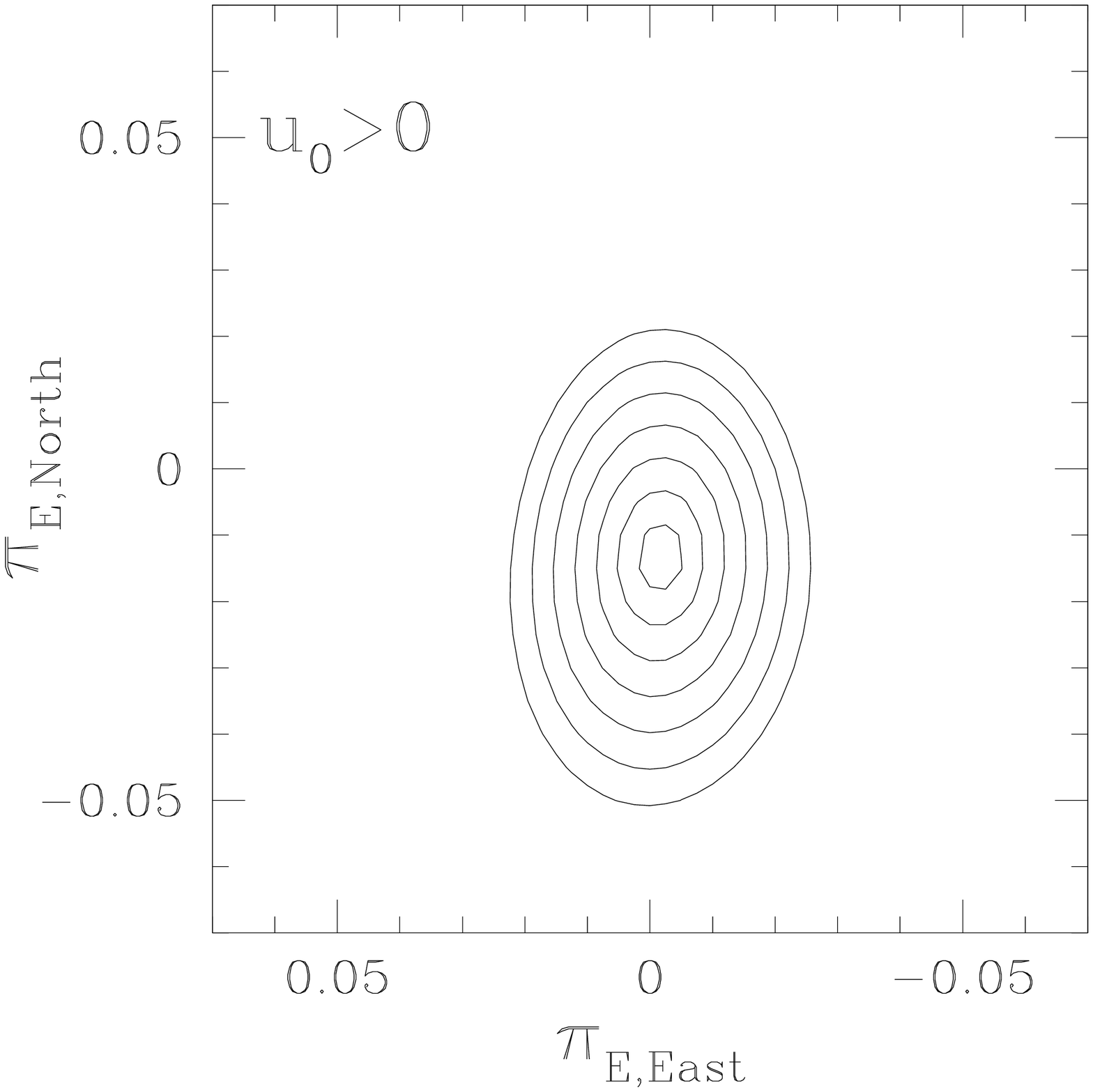}{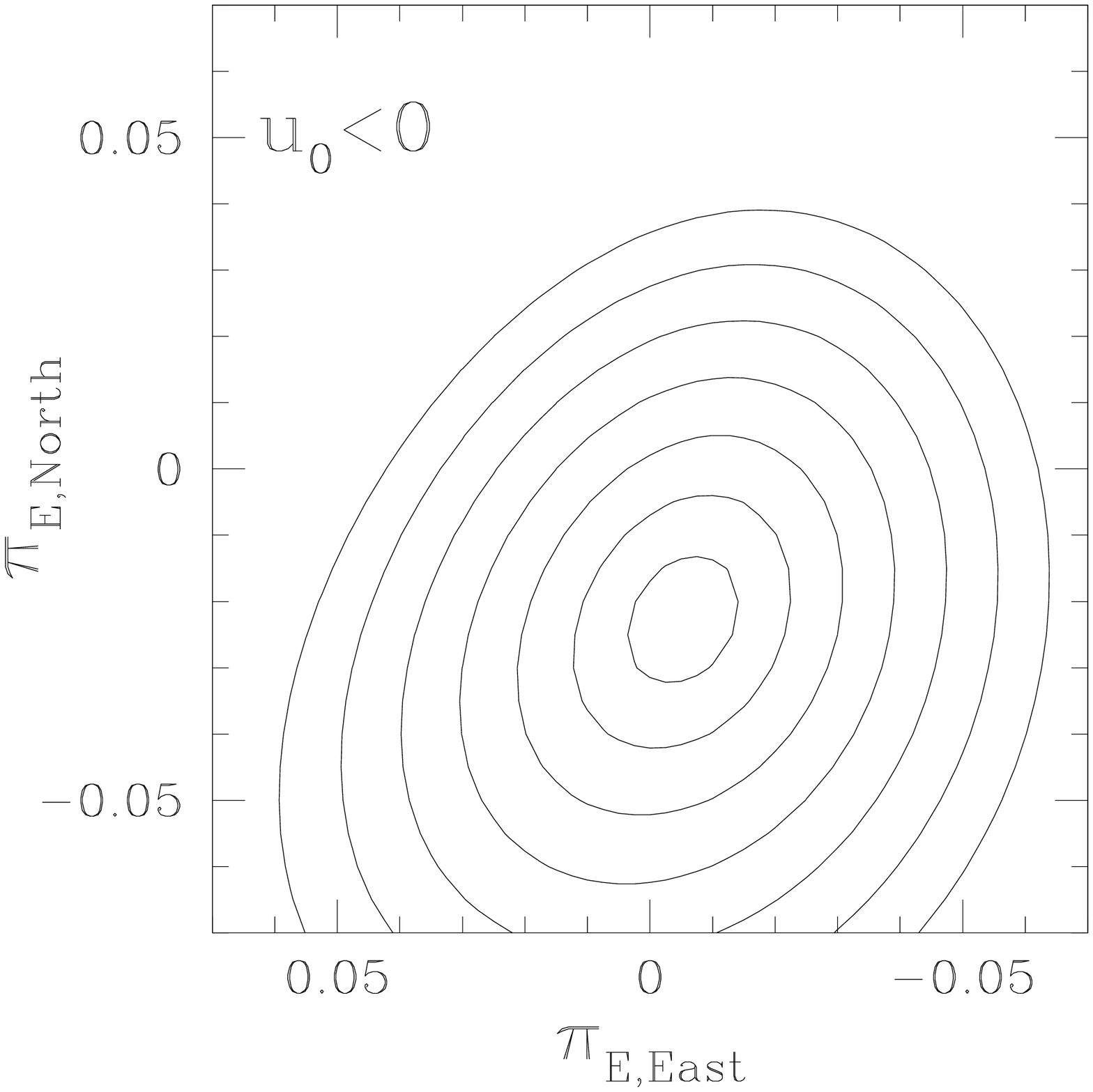}
\caption{$\Delta\chi^2$ contours (1,4,9,16,25,36,49) for the parallax
parameters assuming that 24\% of the light detected in the EROS $I_E$
band is due to blending. Compared to the left contours in Figure
\ref{fg:cont_var_novar}, they are strongly compressed in the
$\pi_{\rme,\perp}$ direction, showing that by this assumption the
degeneracy between $f_{\rm b}$ and $\pi_{\rme,\perp}$ is almost
completely eliminated.}\label{fg:pi_cont_eros}
\end{figure}

Refitting the model described in \S~\ref{sec:params} under this
assumption, we obtain the $\chi^2$ contours of the parallax parameters
for both ($\pm u_0$) solutions shown on Figure
\ref{fg:pi_cont_eros}. Both sets of contours are significantly less
elongated in the $\pi_{\rme,\perp}$ direction when compared to their
analogs in Figure \ref{fg:cont_var_novar}, showing that the degeneracy
is mostly broken, as expected.

\begin{figure}
\plottwo{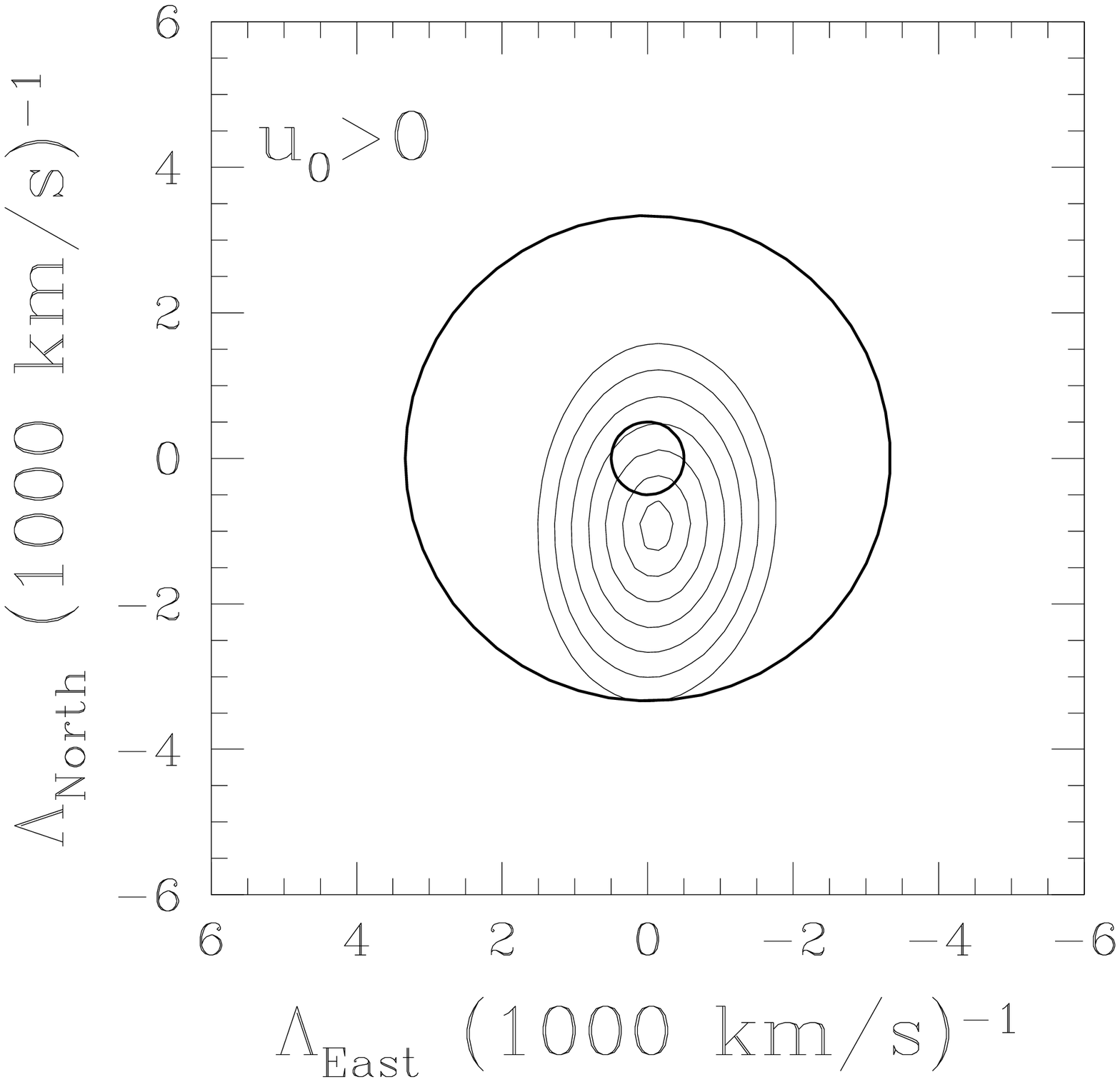}{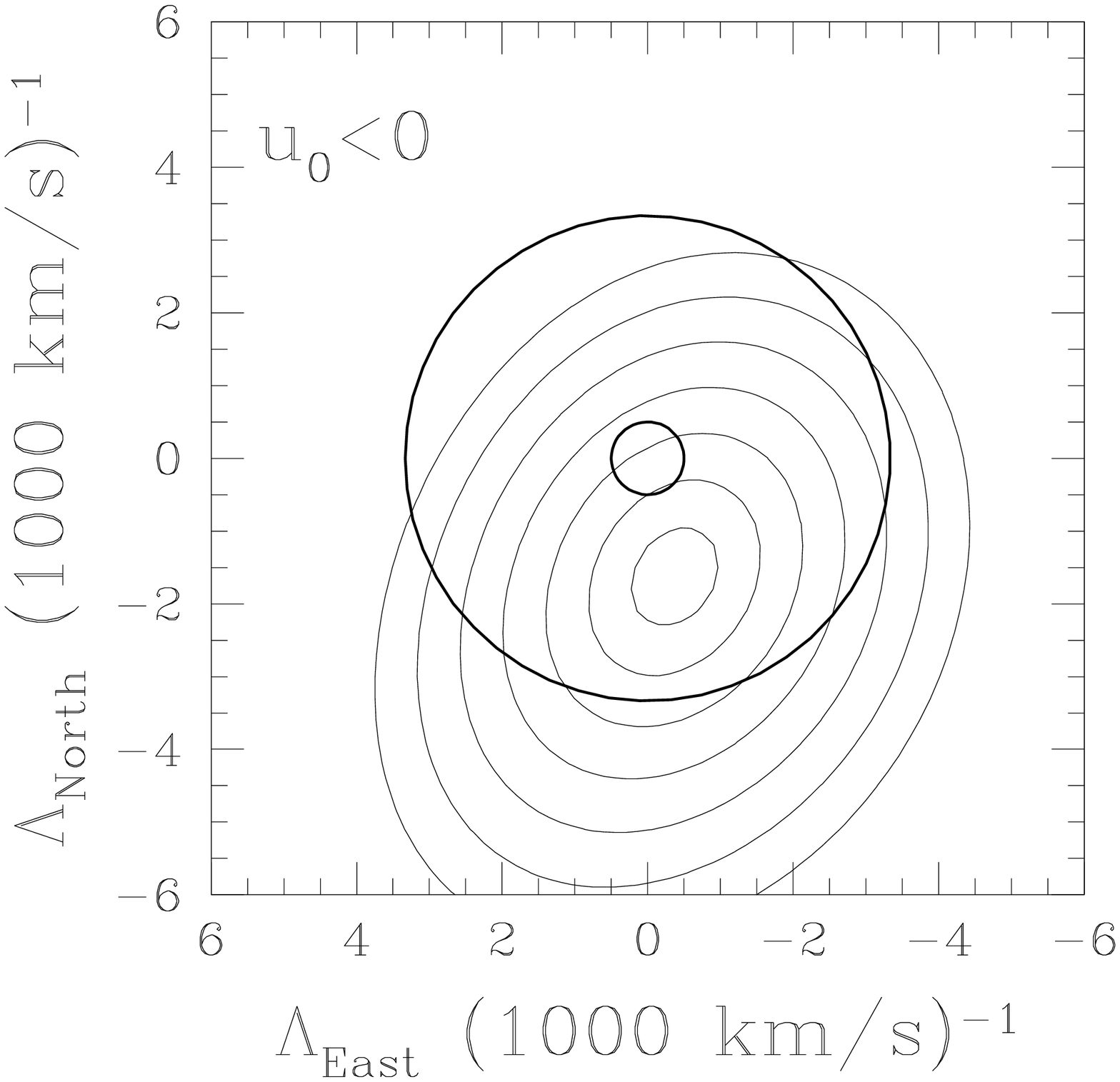}
\caption{$\Delta\chi^2$ contours (1,4,9,16,25,36,49) for $\bLambda$
assuming that 24\% of the light in the EROS $I_E$ band is due to a
blend. As discussed in \S~\ref{sec:eros_fix_blend}, even though the
contours are now nearly circular, the location of the lens cannot be
determined with good confidence.}\label{fg:lambda_cont_eros}
\end{figure}

A better constraint on the microlensing parallax allows for a better
constraint on the projected velocity. Following the analysis detailed
in \S~\ref{sec:location}, we show the $\chi^2$ contours for the
components of vector $\bdv{\Lambda}$ in Figure
\ref{fg:lambda_cont_eros}. These contours are also strongly
compressed, so a better determination of the location of the lens
might be possible.

Employing the statistical test detailed in \S~\ref{sec:location}, we
find that, for the positive $u_0$ solution, SMC self lensing is
favored by factors of 27 and 33 for the larger and smaller SMC assumed
velocity dispersions respectively. However, for the negative $u_0$
solution we find that the factors are 1 and 3, respectively. This
result is very important, since it shows that even if we impose a
strong constraint on the blended light, the location of the lens
cannot be determined with good confidence for this event.

\section{Discussion}\label{sec:discussion}

Variable stars are systematically eliminated from microlensing studies,
but if one of them undergoes a microlensing event, its
variability might actually be useful to constrain the event's nature.

One such an event, MACHO 97-SMC-1, passed the variable-star cuts used
by the MACHO and EROS collaborations. Using several data sets of this
event taken by the MACHO, EROS and OGLE collaborations, we have shown
that if the magnified source star is a regular variable, the parallax
parameter in the direction perpendicular to the projected position of
the Sun, $\pi_{\rme,\perp}$, can be constrained much better than if it
were not a variable: the variability breaks the degeneracy between
$\pi_{\rme,\perp}$ and the blended light. Figure
\ref{fg:cont_var_novar} shows that, if we remove the variability from
both the model and the data, the width of the $\bdv{\pi}_{\rme}$
contours are almost doubled in the $\pi_{\rme,\perp}$ direction while
they remain constant in the $\pi_{\rme,\parallel}$ direction.

We have proposed a statistical test to discriminate between the
possible lens locations (halo or SMC) based on their velocity
distributions and the $\chi^2$ contours of $\bdv{\Lambda}$, the
inverse of the projected velocity. For MACHO 97-SMC-1, the test
revealed that the lens is more likely to be located in the SMC itself
rather than in the Galactic halo but that it is not possible to rule
out the latter location. This test might be useful to discriminate
quantitatively between possible lens locations of other microlensing
events for which a parallax measurement can be performed.

We also showed that, even imposing a strong constraint on the flux,
indicated by OGLE observations, the location of the lens still cannot
be determined with good confidence.

\acknowledgments We wish to thank Alejandro Clocchiatti for providing
us with a spectrum of the MACHO 97-SMC-1 source star.  Work by RJA and
AG was supported by grant AST-0452758 from the NSF.  Support for OGLE
was provided by Polish MEiN grant 2P03D02124, NSF grant AST-0204908
and NASA grant NAG5-12212.  DLW acknowledges research support from the
Natural Sciences and Engineering Research Council of Canada (NSERC) in
the form of a Discovery Grant.

This paper utilizes public domain data obtained by the MACHO Project,
jointly funded by the US Department of Energy through Lawrence
Livermore National Laboratory under contract W7405-ENG-48, the
National Science Foundation through the Center for Particle
Astrophysics of the University of California under cooperative
agreement AST-8809616, and the Mount Stromlo and Siding Springs
Observatory by the Bilateral Science and Technology Program of the
Australian Department of Industry, Technology and Regional
Development.

\appendix

\section{MACHO 97-SMC-1 as an Ellipsoidal
Variable}\label{sec:ellipsoidal}

\citet{udalski97} suggested that the source star is an ellipsoidal
binary system, based on its nearly sinusoidal variations. In such
systems, the binary separation is very small, and either one or both
components are ellipsoidally distorted. Due to the interactions
between them, their rotation and orbital periods are usually
synchronized and the orbits are most likely circularized (see
\citealt{beech85} and references therein). The star appears brighter
at quadrature than at conjunction for two reasons. First, it has a
larger surface area facing the observer. Second, the mean surface
gravity of the exposed area is higher, implying that the surface is
hotter and thus has a higher surface brightness. This leads to
sinusoidal variations with a period equal to half of the orbital
period.

\citet{sahu98} showed that, if MACHO 97-SMC-1 is a binary system, it
is a single-lined spectroscopic one. As outlined by \citet{morris85}
and reviewed and updated by \citet{morris93}, it is possible to
estimate the mass of the secondary star as well as its radius and
other orbital parameters for a single-lined spectroscopic ellipsoidal
binary using its light curve, the mass of the primary and its radial
velocities (for double-lined spectroscopic binaries the mass of the
primary can also be derived). The companion star should be dim enough
to not be detected in the spectrum but massive enough to produce an
ellipsoidal distortion on the primary.

To relate the amplitude of ellipsoidal variations to the binary
parameters, we use equation (6) of \citet{morris85},
\begin{equation}\label{eq:morris_6}
\frac{q R_1^3 \sin^2{i}}{A^3}\ =\ \frac{3.070\ \Delta M_1\
(3-u_1)}{(\tau_1+1)(15+u_1)} , 
\end{equation}
\noindent where $q$ is the binary mass ratio $(m_2/m_1$), $i$ is the
orbital inclination, $R_1$ is the mean radius of the primary, $A$ is
the semi-major axis of the orbit, $\Delta M_1$ is the mean
peak-to-peak amplitude of the primary star alone, $u_1$ is the linear
limb-darkening coefficient for the primary and $\tau_1$ is its gravity
darkening coefficient. These last two parameters ($\tau_1$ and $u_1$)
depend on the wavelength of the observation, $\lambda$. In our
specific evaluation we will assume a $V_E$ filter, but note that the
results are essentially independent of this assumption. Equation (6)
of \citet{morris85} is accurate to first order and will suffice for
this analysis. A more exact version of this equation is given by
\citet{morris93}.

\begin{figure}
\plotone{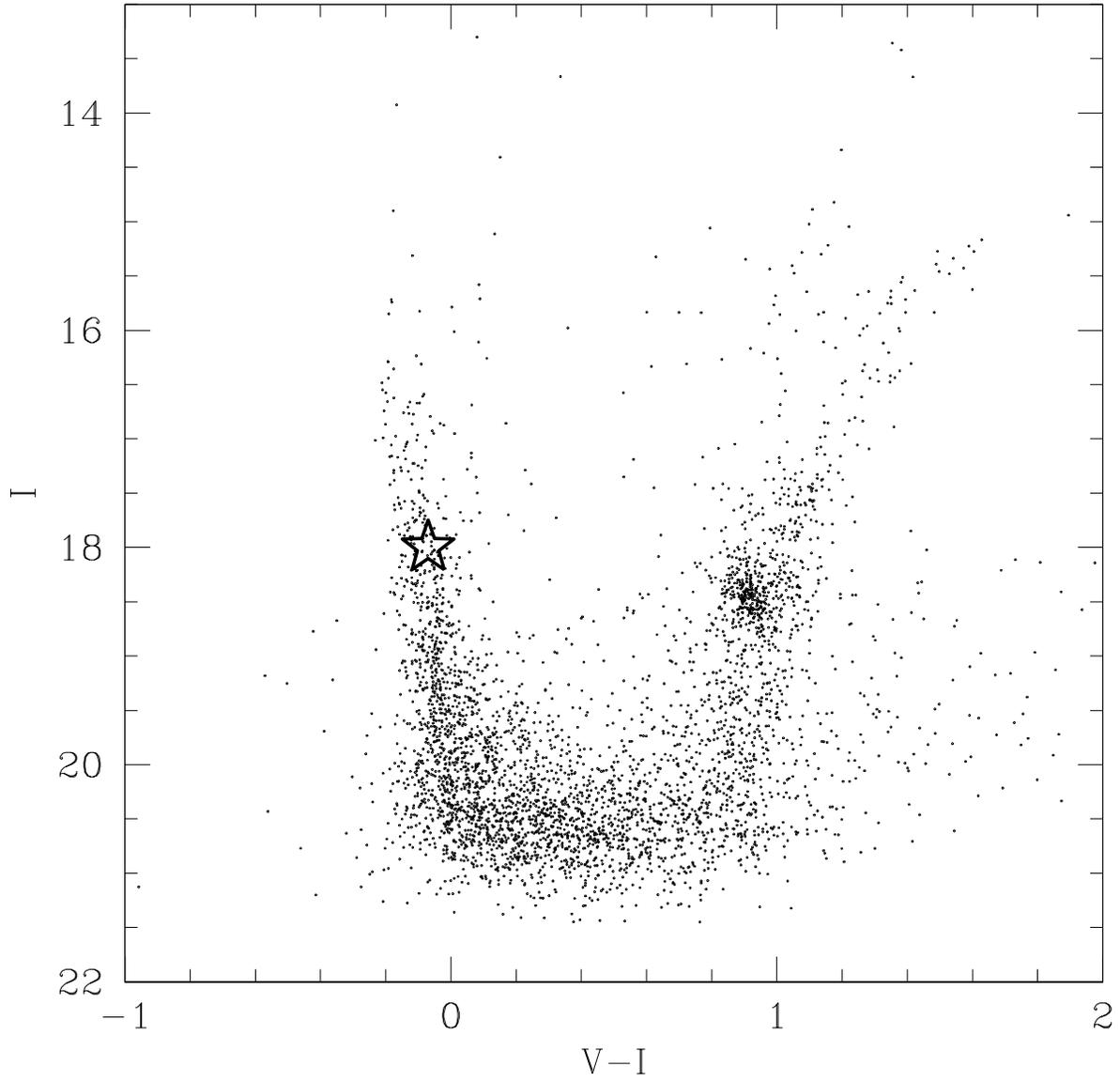}
\caption{Color Magnitude Diagram of the OGLE-II SMC field centered on
MACHO 97-SMC-1. The star marks the position of the source of the MACHO
97-SMC-1 event. This indicates that it is a late B main sequence star,
compatible with the spectroscopic determination of
\citet{sahu98}.}\label{fg:cmd}
\end{figure}

From its position on the color-magnitude diagram (Fig. \ref{fg:cmd})
the source is a late B main sequence star (see also
\citealt{eros98,sahu98}) with an approximate mass of 5
$M_{\odot}$. Using the empirical calibration of \citet{vanbelle99}
between the angular diameter and the $(V-K)$ color yields $R_1\approx
5 R_{\odot}$ if we assume that the value of $(V - K)$ is twice the
value of $(V - I) = -0.05$ and that $D_{\rm SMC} = 60$
kpc. Theoretical models predict a 15\% smaller radius, but this will
not be important for the results of this analysis. The orbital period
$P$ should be twice the variability period, so $P=10.2$ days according
to the variability period found in \S~\ref{sec:results}.

According to the tables of \citet{alnaimiy78}, for the $V_E$ filter
$u_1 \approx 0.35$, while $\tau_1$ is given by equation (10) of
\citet{morris85}
\begin{equation}\label{eq:morris_10}
\tau_1\ =\ \beta \frac{1.43879\times 10^8/\lambda
T_1}{1-\exp(-1.43879\times 10^8/\lambda T_1)}
\end{equation}
\noindent where $T_1$ is the effective temperature of the primary star
($14000$K in this case, \citealt{sahu98}), $\lambda$ is in \AA\ and
$\beta = 0.25$ for early type stars (von Zeipel's Law,
\citealt{vonzeipel24}), and yields $\tau_1 = 0.55$. Using Keplers
Third Law to eliminate $A$ in equation (\ref{eq:morris_6}) and solving
for $\Delta M_1$, we find that
\begin{eqnarray}\label{eq:d_M_1}
\Delta M_1\ =\ 0.0094\sin^2 i\ \left(\frac{R_1}{5 R_{\odot}}\right)^3\
\left(\frac{m_1}{5 M_{\odot}}\right)^{-1}\ \left(\frac{P}{10.2
\rm{days}}\right)^{-2}\ \frac{m_2}{m_1 + m_2}
\end{eqnarray}

From Table \ref{tab:best_fit}, $\Delta M_1 \approx [5/\ln(10)]\
\epsilon_{V_E} = 0.063$ mag for the ``$V_E$'' filter. If we adopt the
most extreme mass parameters, $m_2 \gg m_1$ and $i = 90^{\circ}$, the
right-hand side must increase by a factor of 7 from our estimate to
reproduce the observed $\Delta M_1$.

This type of binary system should also present xallarap distortions on
its microlensing events light curve. In order to test this, we fitted
the data with a standard microlensing model with xallarap and
variability but no parallax and found out that, to the 3-$\sigma$
level, $\chi_{\rme} < 0.015$, where $\chi_{\rme}$ is the xallarap
parameter,
\begin{equation}
\chi_{\rme}\ =\ \frac{a_1}{\hat{r}_{\rme}}.
\end{equation}
That is, $\chi_{\rme}$ is the semi-major axis of the primary
component, $a_1$, in units of the Einstein ring radius projected in
the plane of the source, $\hat{r}_{\rme}$. This small xallarap tends
to reinforce the idea that MACHO 97-SMC-1 is not a binary system and
is therefore not an ellipsoidal variable. However, the small xallarap
might simply be caused by a very large Einstein ring projected on the
source plane, which cannot currently be ruled out since the location
of the lens cannot be determined with good confidence.

\clearpage
\input{tab1}

\input{tab2}

\clearpage

\end{document}

%% file: tab1.tex
\begin{table}
\begin{center}

\begin{tabular}{l l c r  c  r  c  r  c  r}
\hline
Observatory&Filter& &$u_0 > 0$& &$u_0 < 0$& &$u_0 > 0$ & &$u_0 < 0$   \\ 
& & & & & & &NoVar& &NoVar\\
\hline
& & & & & & & & & \\
MACHO      &$V_M$&  &$  306.6$&\ \ &$  306.5$&\ \ &$  305.6$&\ \ &$  305.9$\\
           &$R_M$&  &$ 1057.7$&\ \ &$ 1056.7$&\ \ &$ 1052.0$&\ \ &$ 1053.1$\\
EROS       &$V_E$&  &$  724.5$&\ \ &$  724.5$&\ \ &$  724.5$&\ \ &$  724.6$\\
           &$I_E$&  &$  841.0$&\ \ &$  841.9$&\ \ &$  843.8$&\ \ &$  844.1$\\
MACHO CTIO &$R$&    &$   74.9$&\ \ &$   75.0$&\ \ &$   75.9$&\ \ &$   75.6$\\
OGLE II    &$I$&    &$  315.9$&\ \ &$  315.9$&\ \ &$  316.4$&\ \ &$  316.2$\\
OGLE III   &$I$&    &$  367.3$&\ \ &$  367.3$&\ \ &$  367.3$&\ \ &$  367.3$\\
& & & & & & & & & \\
$\bf Total$& & &$\bf 3688.0$&\ \ &$\bf 3687.8$&\ \ &$\bf 3685.5$&\ \ &$\bf 3686.9$\\
\hline
\end{tabular}
\caption{$\chi^2$ for each filter (for each observatory) for the different models and solutions. NoVar indicates the case for which the best fitted variability of the star has been taken out from the data. Note that, because of round-off, in some cases the $\chi^2$ of the filters/observatories do not add exactly to the total.}\label{tab:chi_2}

\end{center}
\end{table}

%% file: tab2.tex
\begin{table}

\begin{small}
\begin{center}
\begin{tabular}{ c r r c  r r  c  r r  c  r r}
\hline
&\multicolumn{2}{c}{$u_0 > 0$} & &\multicolumn{2}{c}{$u_0 < 0$} & &\multicolumn{2}{c}{$u_0 > 0$ NoVar}& &\multicolumn{2}{c}{$u_0 < 0$  NoVar}  \\ 
& & & & & & & & & & & \\
&Best Fit&Error& &Best Fit&Error & &Best Fit&Error& &Best Fit&Error\\
\hline
$t_0$ (days)            &$1460.041$&$0.338 $&\ \ &$ 1459.936$&$ 0.385 $&\ \ &$1460.014$&$ 0.329$&\ \ &$1459.910$&$ 0.428$\\
$u_0$                   &$   0.419$&$0.048 $&\ \ &$   -0.423$&$ 0.046 $&\ \ &$   0.239$&$ 0.071$&\ \ &$  -0.344$&$ 0.077$\\
$t_{\rme}$ (days)       &$ 124.299$&$9.780 $&\ \ &$  129.059$&$13.689 $&\ \ &$ 188.727$&$44.494$&\ \ &$ 158.647$&$37.382$\\
$\pi_{\rme,N}$          &$  -0.017$&$0.011 $&\ \ &$   -0.028$&$ 0.020 $&\ \ &$  -0.057$&$ 0.015$&\ \ &$  -0.059$&$ 0.032$\\
$\pi_{\rme,E}$          &$  -0.002$&$0.003 $&\ \ &$   -0.005$&$ 0.009 $&\ \ &$  -0.004$&$ 0.003$&\ \ &$  -0.005$&$ 0.008$\\
$\Omega$ (rad/days)     &$   1.226$&$0.000 $&\ \ &$    1.226$&$ 0.000 $&\ \ &$\       $&$\     $&\ \ &$\       $&$\     $\\
$t_{0,\epsilon}$ (days) &$   2.539$&$0.025 $&\ \ &$    2.540$&$ 0.025 $&\ \ &$\       $&$\     $&\ \ &$\       $&$\     $\\
&\multicolumn{11}{c}{MACHO} \\
$f_{s,V_M}$     &$   0.550$&$0.090 $&\ \ &$    0.557$&$ 0.088 $&\ \ &$   0.259$&$ 0.096$&\ \ &$   0.417$&$ 0.128$\\
$f_{b,V_M}$     &$   0.253$&$0.090 $&\ \ &$    0.246$&$ 0.088 $&\ \ &$   0.543$&$ 0.096$&\ \ &$   0.385$&$ 0.127$\\
$\epsilon_{V_M}$&$   0.029$&$0.004 $&\ \ &$    0.028$&$ 0.004 $&\ \ &$\       $&$\     $&\ \ &$\       $&$\     $\\
$f_{s,R_M}$     &$   0.861$&$0.141 $&\ \ &$    0.871$&$ 0.138 $&\ \ &$   0.405$&$ 0.151$&\ \ &$   0.653$&$ 0.200$\\
$f_{b,R_M}$     &$   0.360$&$0.141 $&\ \ &$    0.350$&$ 0.137 $&\ \ &$   0.815$&$ 0.150$&\ \ &$   0.568$&$ 0.200$\\
$\epsilon_{R_M}$&$   0.023$&$0.003 $&\ \ &$    0.023$&$ 0.003 $&\ \ &$\       $&$\     $&\ \ &$\       $&$\     $\\
&\multicolumn{11}{c}{EROS} \\
$f_{s,V_E}$     &$   0.532$&$0.087 $&\ \ &$    0.539$&$ 0.085 $&\ \ &$   0.251$&$ 0.093$&\ \ &$   0.403$&$ 0.124$\\
$f_{b,V_E}$     &$   0.236$&$0.087 $&\ \ &$    0.230$&$ 0.085 $&\ \ &$   0.518$&$ 0.093$&\ \ &$   0.365$&$ 0.124$\\
$\epsilon_{V_E}$&$   0.029$&$0.005 $&\ \ &$    0.029$&$ 0.005 $&\ \ &$   \    $&$ \    $&\ \ &$\       $&$\     $\\
$f_{s,I_E}$     &$   1.226$&$0.201 $&\ \ &$    1.241$&$ 0.196 $&\ \ &$   0.577$&$ 0.215$&\ \ &$   0.929$&$ 0.285$\\
$f_{b,I_E}$     &$   0.487$&$0.201 $&\ \ &$    0.473$&$ 0.196 $&\ \ &$   1.136$&$ 0.215$&\ \ &$   0.784$&$ 0.285$\\
$\epsilon_{I_E}$&$   0.024$&$0.004 $&\ \ &$    0.024$&$ 0.004 $&\ \ &$\       $&$\     $&\ \ &$\       $&$\     $\\
&\multicolumn{11}{c}{MACHO CTIO} \\                                   
$f_{s,R}$       &$  24.133$&$4.126 $&\ \ &$   24.426$&$ 3.950 $&\ \ &$  11.384$&$ 4.134$&\ \ &$  18.351$&$ 5.581$\\
$f_{b,R}$       &$   7.277$&$4.155 $&\ \ &$    6.965$&$ 3.967 $&\ \ &$  20.026$&$ 4.101$&\ \ &$  13.032$&$ 5.574$\\
$\epsilon_{R}$  &$   0.030$&$0.005 $&\ \ &$    0.029$&$ 0.005 $&\ \ &$\       $&$\     $&\ \ &$\       $&$      $\\
&\multicolumn{11}{c}{OGLE II} \\                                       
$f_{s,I}$       &$   0.734$&$0.128 $&\ \ &$    0.737$&$ 0.127 $&\ \ &$   0.347$&$ 0.134$&\ \ &$   0.551$&$ 0.174$\\
$f_{b,I}$       &$   0.204$&$0.129 $&\ \ &$    0.201$&$ 0.127 $&\ \ &$   0.590$&$ 0.133$&\ \ &$   0.386$&$ 0.174$\\
$\epsilon_{I}$  &$   0.031$&$0.006 $&\ \ &$    0.031$&$ 0.006 $&\ \ &$\       $&$\     $&\ \ &$\       $&$\     $\\
&\multicolumn{11}{c}{OGLE III} \\                                       
$f_{s,I}$       &$   0.684$&$0.001 $&\ \ &$    0.684$&$ 0.001 $&\ \ &$   0.684$&$ 0.001$&\ \ &$   0.684$&$ 0.001$\\
$f_{b,I}$       &$   0.254$&$      $&\ \ &$    0.254$&$       $&\ \ &$   0.254$&$      $&\ \ &$   0.254$&$      $\\
$\epsilon_{I}$  &$   0.033$&$0.003 $&\ \ &$    0.033$&$ 0.003 $&\ \ &$\       $&$\     $&\ \ &$\       $&$\     $\\

\hline
\end{tabular}
\end{center}
\end{small}

\caption{Best fit parameters with their respective errors for the different models and solutions. NoVar indicates the case for which the best fitted variability of the star has been taken out from the data. $t_{0\epsilon}$ is defined as $\phi/\Omega$, where $\phi$ is the phase defined in equation (\ref{eq:f_t_2}). The time of the maximum, $t_0$, is measured in $JD-2449000$, while the fluxes are measured in arbitrary units. The seemingly excessive number of decimal places included for each parameter are potentially important to reproduce these results, because the correlations among some parameters are high.}\label{tab:best_fit}

\end{table}